\documentclass{elsart}

\usepackage{graphicx}
\usepackage{array}
\usepackage{dcolumn}
\usepackage{amssymb}
\usepackage{ctable}
\usepackage{rotating}

\newcommand{\mue}{\mu_e}
\newcommand{\gcmic}{g/cm$^{3}$} 

\newcommand{\updown}[2]{$_{#2}^{#1}$}

\newcommand{\avgrate}{\langle\lambda_{\rm ec}\rangle}
\newcommand{\avgrateNS}{\langle\lambda_{\rm ec}^{\rm ns}\rangle}

\newcommand{\nuprob}{{\cal N}^{\nu}(\varepsilon_\nu)}
\newcommand{\nuprobi}{{\cal N}_i^{\nu}(\varepsilon_\nu)}

\newcommand{\avgdistr}{\langle {\cal N}_{\nu}(E_\nu) \rangle}

\newcommand{\TFD}{T_{\rm FD}}
\newcommand{\lambdaEC}{\lambda_{\rm ec}}
\newcommand{\sigmaEC}{\sigma_{\rm ec}}
\newcommand{\Msol}{M_{\odot}}
\newcommand{\kB}{k_{\rm B}}
\newcommand{\Qif}{Q_{i\!f}}

\newcommand{\secinv}{sec$^{-1}$}
\newcommand{\beq}{\begin{equation}}
\newcommand{\eeq}{\end{equation}}

\newcommand{\refeq}[1]{(\ref{#1})}
\newcommand{\widebar}{\bar}
\newcommand{\MC}{\multicolumn}

\newcommand{\simlt}{\mathrel{\lesssim}}

\newcolumntype{d}[0]{D{.}{.}{-1}}

\newlength{\figsize}
\newlength{\figsizeBig}
\begin{document}

\begin{frontmatter}
\title{Improved estimate of electron capture rates on nuclei 
during stellar core collapse}

\author[ITPA]{A.\ Juodagalvis}
\author[GSI,TU,Frankfurt]{K.\ Langanke}
\author[ORNL]{W.R.\ Hix}
\author[GSI]{G.\ Mart\'{\i}nez-Pinedo}
\author[CFNUL]{J.M.\ Sampaio}

\address[ITPA]{VU ITPA,
A.\ Go{\v s}tauto St.\ 12, 01108 Vilnius, Lithuania}
\address[GSI]{GSI Helmholtzzentrum f\"ur Schwerionenforschung, 
Planckstr.\ 1, 64291 Darmstadt, Germany}
\address[TU]{Institut f\"ur Kernphysik, TU Darmstadt,
Schlossgartenstr.\ 9, 64291 Darmstadt, Germany}
\address[Frankfurt]{Frankfurt Institute for Advanced Studies, 
Ruth-Moufang-Str.\ 1, 60438 Frankfurt, Germany}
\address[ORNL]{Physics Division, ORNL, 
  P.O.\ Box 2008, Oak Ridge, TN 37831-6373, USA}
\address[CFNUL]{CFNUL, Av.\ Prof.\ Gama Pinto 2, 1649-003 Lisboa, Portugal}

\date{July 12, 2010}
\begin{abstract}
Electron captures on nuclei play an important role
in the dynamics of the collapsing core of a massive star that 
leads to a supernova explosion. Recent calculations
of these capture rates were based on microscopic models which
account for relevant degrees of freedom. Due to computational
restrictions such calculations were limited to a modest
number of nuclei, mainly in the mass range $A=45$--$110$.
Recent supernova simulations show that this pool of nuclei,
however, omits the very neutron-rich and heavy nuclei
which dominate the nuclear composition during the
last phase of the collapse before neutrino trapping.
Assuming that the composition is given by 
Nuclear Statistical Equilibrium
we present here electron capture rates for collapse 
conditions derived from individual rates for roughly $2700$ individual
nuclei. For those nuclei which dominate in the early stage of the collapse,
the individual rates are derived within the framework of microscopic models,
while for the nuclei which dominate at high densities we have derived the rates
based on the Random Phase Approximation with a global parametrization
of the single particle occupation numbers. In addition, we have improved previous
rate evaluations by properly including screening corrections to the reaction rates into account.
\end{abstract}

\end{frontmatter}
\maketitle

\section{Introduction}
Near the end of their lives, the core of a massive star ($M\geq
10\Msol$) consists predominantly of iron and its nuclear neighbors,
the ``iron group'' elements.  Since nuclear burning processes continue
in the layers above, this iron core grows, approaching the
Chandrasekhar mass limit, and begins contracting at an ever increasing
rate.  In the inner regions of the core, this collapse is subsonic and
homologous, while the outer regions collapse supersonically. With
densities larger than $10^9$ \gcmic, both transport of electromagnetic
radiation and heat conduction are very slow compared with the
time-scale of the collapse. The majority of the energy is transported
away from the core by neutrinos, originating mostly from electron
captures on protons and nuclei \cite{Bethe90}. Escaping neutrinos
reduce the entropy of the core, forcing nucleons to remain as part of
nuclei, until the neutrinos become trapped at densities
$10^{11}$--$10^{12}$~\gcmic.  $\beta^-$-decay processes are
effectively blocked by the electron degeneracy, causing nuclei in the
core to become progressively more neutron-rich.

Once the density in the innermost part of the collapsing star exceeds
that of nuclear matter, the collapse is halted by the short-range repulsion
of the nuclear interaction.  The supersonically infalling matter
of the outer core bounces off this extremely stiff inner part,
reversing its velocity and forming a shock wave that propagates outwards
through layers of lower density.  The shock weakens as it loses energy to nuclear dissociation of the shocked matter and once the shock reaches the neutrino emission surface (or neutrinosphere), the remaining energy of the shock is carried away by escaping neutrinos. This causes the shock to stall and become an accretion shock before it can
drive off the envelope of the star
\cite{Bruenn85,Bruenn89,Rampp00,Mezzacappa01,Liebendorfer01}.

In the present supernova paradigm, the intense neutrino flux emerging
from the proto-neutron star heats the matter just behind the stalled
shock, eventually reenergizing it sufficiently to drive off the
envelope and produce a supernova explosion
\cite{Wilson85,BetheWilson85}.  Unfortunately, despite significant
progress in supernovae modeling, complete self-consistent simulations
often fail to produce explosions
\cite{Mezzacappa01,Liebendorfer01,Buras03,Liebendorfer05,Buras06}. 
(However, recent results have been more promising
\cite{Burrows.Livne.ea:2006,Marek.Janka:2009,Janka.Langanke.ea:2007,
Bruenn09}.)
One potential improvement to these models is replacement of incomplete
or inaccurate treatments of the wide variety of nuclear and weak
interaction physics that are important in the supernova mechanism. In
particular, electron captures play a dominant role during the collapse
as they significantly alter the lepton fraction and entropy of the
inner core.  These quantities, in turn, determine the structure of the
core, and the strength and location of the initial supernova shock. As
a result, the treatment of electron captures significantly influences
the initial conditions for the entire post-bounce evolution of the
supernova.

Following the pioneering work of Fuller, Fowler and
Newman~\cite{Fuller.Fowler.Newman:1980,%
Fuller.Fowler.Newman:1982b,Fuller.Fowler.Newman:1982a,%
Fuller.Fowler.Newman:1985}, extended sets of electron capture rates
have been calculated for heavy nuclei based on rather sophisticated
nuclear structure models and advanced computer algorithms. The rates
of nuclei in the mass range of $A=45$--$65$ (around $100$ nuclides;
hereafter the LMP pool) are derived from large-scale shell model
diagonalizations~\cite{Caurier.Langanke.ea:1999,%
Langanke.Martinez-Pinedo:2000,Langanke.Martinez-Pinedo:2001}
performed in the complete $pf$ shell at a truncation level which
guarantees the virtual convergence of the level spectra and the
Gamow-Teller (GT) strength
distributions. In~\cite{Caurier.Langanke.ea:1999,%
Langanke.Martinez-Pinedo:2000,Langanke.Martinez-Pinedo:2001}
the rates for electron captures (and other weak processes) were
derived solely on the basis of allowed transitions as contributions
from forbidden transitions can be neglected for those stellar
conditions (presupernova evolution) where nuclei in the mass range
$A=45$--$65$ dominate the matter composition.

Electron capture rates for about $80$ heavier nuclei in the mass range
of $A=66$--$112$ have been derived within the framework of a hybrid
model \cite{LMS} (hereafter, the LMS pool), with the rates calculated
via the Random Phase Approximation (RPA) based on average thermal
nuclear states characterized by occupation numbers calculated within
the Shell Model Monte-Carlo (SMMC) approach \cite{Lang03,SMMC97}. The
SMMC model considers relevant finite-temperature effects and
correlations among nucleons \cite{LKD01} which both have been
identified as important for the description of stellar electron
capture rates for heavy nuclei \cite{Lang03}.  Allowed (i.e.\ GT) and
forbidden transitions have been considered in the rate calculations of
Ref.\ \cite{Lang03}.

The consequences of the LMP rates in the presupernova evolution of a
massive star were studied by Heger {\em et al} \cite{Heger01a,Heger01b}. 
These rates have also been employed to study thermonuclear supernovae 
(see, e.g., \cite{Calder07,Brachwitz00}). 
The effects of improved nuclear electron
capture during core collapse was studied in Refs.\
\cite{Langanke03,Hix03}, where the LMSH tabulation 
was constructed by folding the LMP and LMS rates (supplemented by the
rates of Fuller, Fowler and Newman for lighter nuclei with mass
numbers $A<45$) with a detailed calculation of the nuclear composition
assuming Nuclear Statistical Equilibrium (NSE). The assumption of NSE
among free nucleons and discrete nuclei
is well justified during core collapse until densities of order
$10^{13}$~\gcmic\ are reached.

The use of the LMP+LMS electron capture rates in supernova simulations
for a wide range of relevant collapse conditions
\cite{Langanke03,Hix03} showed the importance of a correct treatment
of these nuclear processes.  Previous simulations assumed that
captures on nuclei with neutrons above the $pf$-shell closure ($N>40$)
were negligible due to Pauli blocking of the GT transitions.
(Calculations of thermal unblocking of the GT transitions were
reported in \cite{Fuller82,Cooperstein84}, but found rather small effects at
densities before neutrino trapping. Very recently first attempts have been 
reported to derive the stellar electron capture rates consistently 
based on the finite temperature Random Phase Approximation
\cite{Paar09} or the thermofield dynamics formalism \cite{Dzhioev10}. Comparison
to the SMMC results indicate that the GT unquenching across the shell gap 
requires higher-order correlations than currently accounted for in 
these approaches \cite{Dzhioev10}.)  In these simulations, electron
captures on free protons dominated.  However, thermal excitation of
nuclei and, more importantly, correlations among the nucleons can
effectively unblock the GT transitions in neutron-rich nuclei,
resulting in rates which are several orders of magnitude larger than
assumed before \cite{LKD01}.  In addition, although the captures on
protons remain faster per particle, the total {\em reaction\/} rate
for heavy nuclei is greater than for free protons under the collapse
conditions \cite{Langanke03,Hix03} due to the large abundances of
heavy nuclei.  As a further result, the spectra of neutrinos emitted
by the electron capture process are significantly changed as the heavy
neutron-rich nuclei have noticeably larger $Q$ values than protons.

An important consequence of the dominance of electron captures on
nuclei over free protons is the faster decrease of the lepton fraction
at high densities ($\rho\gtrsim 10^{11}$~\gcmic) and 
temperatures ($T\approx1$~MeV) as under these conditions the nuclear
composition is dominated by nuclei for which electron captures were
originally neglected.
Once neutrino trapping sets in at densities around 
$\rho\sim 10^{12}$~\gcmic\ the deleptonization is hindered by
final-state neutrino
blocking. Recently inelastic neutrino-nucleus neutral current
reactions have been found to have very small effects on the supernova
dynamics, but a larger effect on the high energy tail of the neutrino
spectrum \cite{Langanke08}.

In spite of a smaller inner lepton fraction, resulting in a smaller
initial proto-neutron star and a weaker shock, the shock
ultimately travels slightly further out before it stalls in
simulations which consider electron captures on nuclei \cite{Hix03}.
The captures on nuclei from the LMP pool dominate in the outer regions
of the core.  Since these rates are in general smaller than the
previously used rates from Fuller, Fowler and Newman
~\cite{Fuller.Fowler.Newman:1980,%
  Fuller.Fowler.Newman:1982b,Fuller.Fowler.Newman:1982a,%
  Fuller.Fowler.Newman:1985}, the
deleptonization in the outer regions is smaller. The larger electron
fraction in the outer layers of the core increases the degeneracy
pressure and slows the collapse. A slower collapse reduces the rate at
which density increases and, hence, the ram pressure that opposes the
shock.  In spherically symmetric simulations, this more than
counteracts the effects of the weaker, deeper initial supernova shock.

The calculation of supernova-relevant electron capture rates requires
an appropriate nuclear model to determine the individual capture rates
as well as a reliable account of the many nuclei present in the matter
composition and of their individual abundances. A shortcoming of the
LMSH tabulation was the decreasing fraction of the nuclear
composition included in the LMP+LMS set of electron capture rates as
collapse continues.  In fact at the densities around neutrino trapping
the nuclear composition becomes dominated by nuclei heavier than those
considered in this set.  To minimize this problem, the LMSH
tabulation derives an average electron capture rate per heavy
nucleus on the basis of those nuclei for which individual rates are
available. The total rate is calculated from this average rate and
the heavy nucleus abundance provided by the equation of state.
However, this average becomes increasingly dominated by a few nuclei,
normally the heavier, making it
increasingly uncertain.  It was suggested \cite{Langanke03} to use a
parameterized electron capture rate for the missing nuclei at higher
densities. However, a recent study revealed that the
parameterization is too simple and cannot be used at all conditions
during the collapse \cite{Juodagalvis08}.

This paper presents electron capture rates for a pool of nuclei which
has been enlarged in two ways. The first addition is achieved by
calculating the capture rates within the hybrid model as used in
\cite{Langanke03} for $170$ additional nuclei.  The SMMC+RPA pool
(SMMC pool, below) now covers about $250$ nuclei in the mass range of
$A=66$--$120$ and the proton number range of $Z=28$--$45$.  As the SMMC
calculations are rather time consuming, it is prohibitive to perform
the required studies for the many nuclei which are present in the
latest stages of the collapse.  This prompts a new approach to expand
the pool further, using the fractional occupation numbers of the
various shells in the ground state of the parent nucleus as calculated
from a Fermi-Dirac (FD) parameterization in place of the SMMC
results.  The reduced cost of this approach allows the inclusion of
more than $2200$ additional nuclei to the pool.  To improve this
approach, the parameters of the Fermi-Dirac distribution were adjusted
so that the electron-capture cross sections matched those available
for the SMMC pool of nuclei.
The $Q$ values needed to calculate the cross sections were derived
from the predicted masses of the finite-range droplet model
\cite{Moller95}, if they were not available from Audi {\em et
  al\/} compilation \cite{Audi03}.  The nuclei added in this way are
in the range of proton and neutron numbers $Z=28$--$70$ and
$N=40$--$160$. We will refer to them as the FD+RPA pool.  We further
include the FFN rates~\cite{Fuller.Fowler.Newman:1980,%
  Fuller.Fowler.Newman:1982b,Fuller.Fowler.Newman:1982a,%
  Fuller.Fowler.Newman:1985}, for nuclei with mass numbers $A<45$.
Since the majority of these $100$ nuclei are from the $sd$ shell, we
refer to them as the $sd$ pool.

\begin{figure*}[tbp]
\begin{center}
\includegraphics*[width=\figsizeBig, 
angle=0]{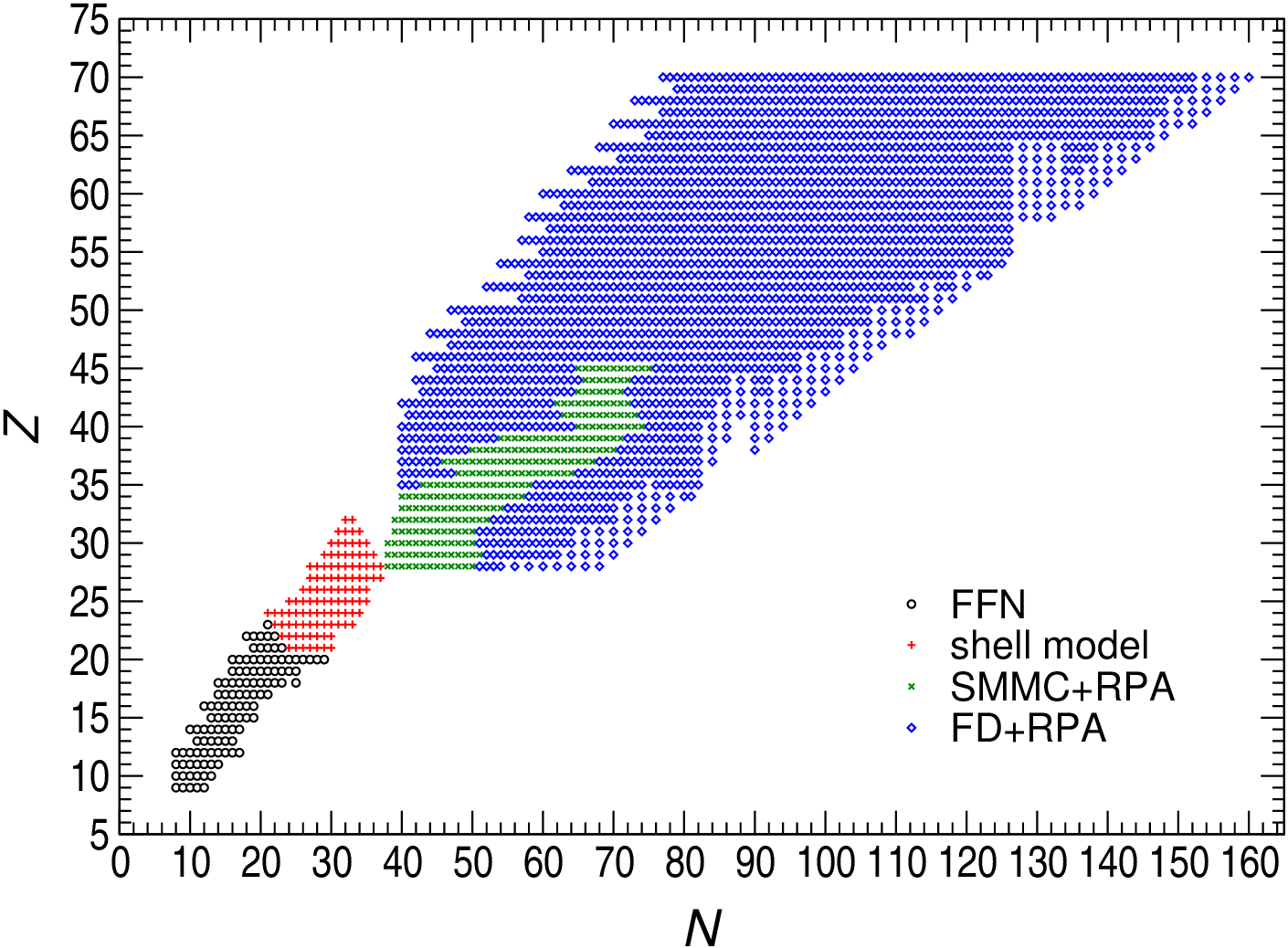}
\caption{\label{fig-nuclchart} (Color online) 
  Nuclei included in the calculation of the NSE-averaged rates and spectra.
  The $sd$ pool is marked by circles, the shell model pool is marked by
  plusses, the SMMC+RPA pool is marked by crosses, and the FD+RPA pool
  is marked by diamonds.}
\end{center}
\end{figure*}

To have an overview of the pools, Fig.\ \ref{fig-nuclchart} shows the
nuclei included in our rate evaluation distinguishing the method used
for the individual rates. With four approaches all contributing,
validation is an important consideration. Ideally, one would validate
the various approaches against experimental data. Unfortunately, such a
procedure is severely limited by the lack of data for
excited states, while the relative weight of the ground state in the
thermal ensemble at temperatures present in the collapsing core is
rapidly decreasing. Furthermore, the important role played by the
phase space has to be recognized as well.  The electron chemical
potential grows significantly faster than the average $Q$-values of
nuclei present in the core. As a consequence, the stellar electron
capture rates are sensitive to detailed Gamow-Teller distributions
only at low densities. At higher densities it suffices that the total
GT strength and its energy centroid are well described. This calls for
more elaborate nuclear models to be used to derive the capture rates
for nuclei at low densities than is required for the rates needed at
larger densities. We have considered these facts in our strategy for
the validation process. At low densities, which is also at low temperatures 
when nuclei are present, the ground states contribute relatively strongly to the
capture rates. For such conditions the core composition is dominated
by nuclei with the mass numbers $A=45$--$65$ for which the shell model
diagonalization calculation can be performed. Such shell model rates
are adopted here. It has been proven by extensive comparison to experiment that the
shell model describes measured $GT_+$ strength distributions in
this mass range very well and also gives a very good account of the
spectra at low excitation energies
\cite{Caurier.Langanke.ea:1999,Langanke.Martinez-Pinedo:2000,Langanke.Martinez-Pinedo:2001,Caurier05}.  
By the time nuclei with $A>65$, for which shell model diagonalization
calculation are prohibitive due to computational
restrictions, dominate the core composition, the density, and
accordingly the electron chemical potential, has grown sufficiently
that the capture rates are mainly sensitive to the total GT strength
and its centroid. For such nuclei we have performed RPA calculations
with the occupation numbers determined from SMMC calculations or from
the simple FD parametrization. This procedure is validated by comparing the
shell model and SMMC+RPA results for selected nuclei, indeed finding
good agreement, if the electron chemical potential is appropriately large. 
Finally we derive a simple parametrization of the SMMC
occupation numbers and show that this FD parametrization reproduces the
SMMC+RPA capture rates quite well. Such a comparison of the rates in
the overlapping regions together with the outlined reasoning on the
rate sensitivity to the conditions gives us confidence that the
approach used is valid.

Having produced the individual rates, we have derived the NSE-averaged
electron capture rates for the appropriate conditions during the
collapse based on the pool of nearly $2700$ nuclei.  An extended
tabulation of the capture rates and the corresponding neutrino
spectra is available in electronic form upon request from the
authors for a wide grid of stellar conditions (defined by temperature,
the matter density and the electron-to-baryon ratio).  This tabulation is 
most appropriate for the study of core collapse supernovae, where the large 
range of density and the neutrino spectral information are necessary.  This 
tabulation is also appropriate for use in thermonuclear supernovae, where 
the range of density and electron fraction are a subset of those occurring 
in core collapse supernovae, however the value of the wider nuclear range 
and neutrino spectral data considered here is lost.  Thus for the thermonuclear supernova problem, this tabulation is essentially equivalent to that of 
Seitenzahl et al.\ \cite{Seitenzahl09}, which also folds LMP reactions over 
an NSE abundance distribution.  This tabulation is wholly inappropriate for 
the study of electron capture in X-ray burst ashes (see, e.g.\ \cite{Gupta07}), 
because the ash temperature is insufficient to justify the use of NSE.

The paper is organized as follows. Section \ref{sect-Model} describes
the hybrid model and some computational details. Screening corrections
to the rates are also introduced there, but the corresponding
formalism is presented in an appendix.  Section \ref{sect-Results}
discusses the obtained results. We conclude with section
\ref{sect-Conclusion}.

\section{Theoretical model}
\label{sect-Model}

The hybrid SMMC+RPA model was proposed in \cite{LKD01} to compute
electron capture rates on nuclei which required such large model
spaces for which diagonalization shell model calculations are not
yet feasible.  The hybrid model is computationally feasible, but
simultaneously incorporates relevant nuclear structure physics as
configuration mixing (caused by nucleon correlations) and thermal
effects.  A pairing+quadrupole residual interaction~\cite{PPQQ} was
used which avoids the sign problem associated with using realistic
interactions in SMMC studies~\cite{SMMC97}.  SMMC calculations at
finite temperature are used to obtain occupation numbers for the
various neutron and proton valence shells in the parent nucleus, which
are then used to calculate electron capture cross sections and rates
within a Random Phase Approximation (RPA) approach with partial shell
occupancies (the method is explained in Ref.\ \cite{KLV99}).

We have performed hybrid SMMC+RPA calculations of electron capture
rates for additional nuclei extending the pool of nuclei used in
\cite{LMS,Langanke03} to more neutron-rich nuclei and filling
occasional holes in isotopic chains.  As mentioned above, the first
step in the hybrid approach is to obtain the shell occupation numbers
for protons and neutrons using the SMMC.  Using the same interaction
and model space as in \cite{Langanke03} (i.e.\ full $pf$-$gds$ shells
with $50$ valence orbitals for protons and neutrons), we have
calculated such occupation numbers at finite temperatures for $168$
nuclei with $N\leq61$. For nuclei with even larger neutron number we
switched to the $f_{5/2}p$-$gds$-$h_{11/2}$ model space, changing the
pairing and quadrupole strength parameters of the residual interaction
to account for the change in the model space.  These parameters were
adjusted such as both model spaces predict similar properties for
selected nuclei.  The single-particle energies were taken from
\cite{LDN06}.  The energy of the $h_{11/2}$ was determined using the
same Woods-Saxon parameters as in~\cite{LDN06} placing it at
$15.12$~MeV above the $f_{7/2}$ shell.

The radial wave functions as well as single-particle energies for the
RPA calculation (i.e.\ the second step for the hybrid method)
were taken from the Woods-Saxon potential:
\begin{eqnarray}
  V(r) &= & V_0\ \left[ 1 + \exp\left\{\frac{r-R}{a}\right\}\right]^{-1},
  \nonumber \\
  V_{LS} &=& \frac12\lambda \left(\frac{\hbar}{Mc}\right)^2\, \frac1r
    \frac{\textrm{d}V(r)}{\textrm{d}r}\, {\vec \ell \cdot \vec s},
   \\
  V_{Coul} &=& (Z-1)\,\frac{e^2}{r}\, \times\ 
  \left\{
  \begin{array}{ll}
     \frac32\frac{r}{R}  - \left(\frac{r}{R}\right)^3 &,\ r<R \\
     1 & ,\ r\geq R
  \end{array}
     \right.
   \nonumber
\end{eqnarray}
The depth of the potential
was adjusted to reproduce the proton separation energy in the
parent nucleus and the neutron separation energy in the daughter
nucleus, as discussed in~\cite{Cooperstein84}. 
Other parameters of the Woods-Saxon potential
were: $r_0=1.27$~fm, $R=r_0 A^{1/3}$, $a=0.65$~fm, $\lambda=32.$
The separation energies were calculated
using masses from the Audi {\em et al} compilation \cite{Audi03},
supplemented by predictions from the finite-range droplet model
\cite{Moller95}. The same masses were used to evaluate the $Q$ value
for the calculation of electron capture cross sections as will be
discussed later.

Following the spirit of RPA+BCS calculations we have shifted the
energies of the unoccupied proton states by the amount of the pairing
energy estimated as $12/\sqrt{A}$~MeV to avoid the appearance of
spurious zero energy transitions in situations where orbits are
partially filled. It should be noted that this energy shift does not
affect the physical transitions that take place between occupied
proton states and unoccupied neutron states. Ref.~\cite{Langanke03}
has used a constant energy shift of 2.5~MeV which leads to slightly
larger capture rates. For consistency we have repeated the
calculations of the capture rates for the LMS pool of nuclei with the
$A$-dependent energy shift.

In the RPA calculations we used the Landau-Migdal force as the
residual interaction with the parameters taken
from~\cite{MigdalForce}, except for the overall scale factor
$C_0=302$~MeV~fm$^3$~\cite{MigdalC0}.  In the calculation of the rates
we included all multipole transitions with $J\leq3$. The dependence of
the operators on momentum transfer has been considered which leads to
a reduction of the capture cross sections at high electron
energies~\cite{LMS}. Motivated by global RPA calculations for muon
capture on nuclei~\cite{muonc1,muonc2}, the $J=1^+$ multipole
transitions have been quenched by a factor of $(0.7)^2$ but not the
other.

\begin{figure}[tbp]
\begin{center}
\includegraphics*[width=\figsize,
angle=0]{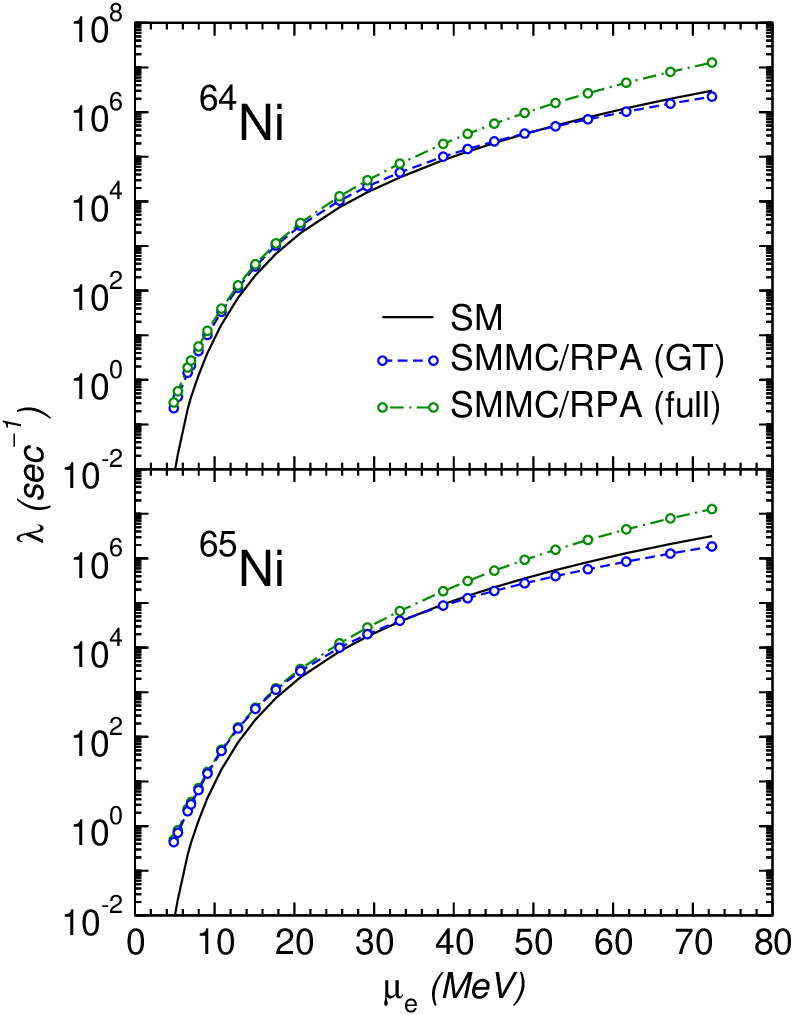} 
\caption{\label{fig-NiIsotopes} (Color online) A comparison of the
  electron capture rates on $^{64,65}$Ni calculated from the
  diagonalization shell model (only allowed contributions) and the
  hybrid SMMC+RPA model (both allowed and forbidden
  contributions). Stellar conditions of the $25\Msol$ trajectory
  (see Table \ref{tbl-Zones}) are used.  }
\end{center}
\end{figure}
\begin{figure}[tbp]
\begin{center}
\includegraphics*[width=\figsize, angle=0]{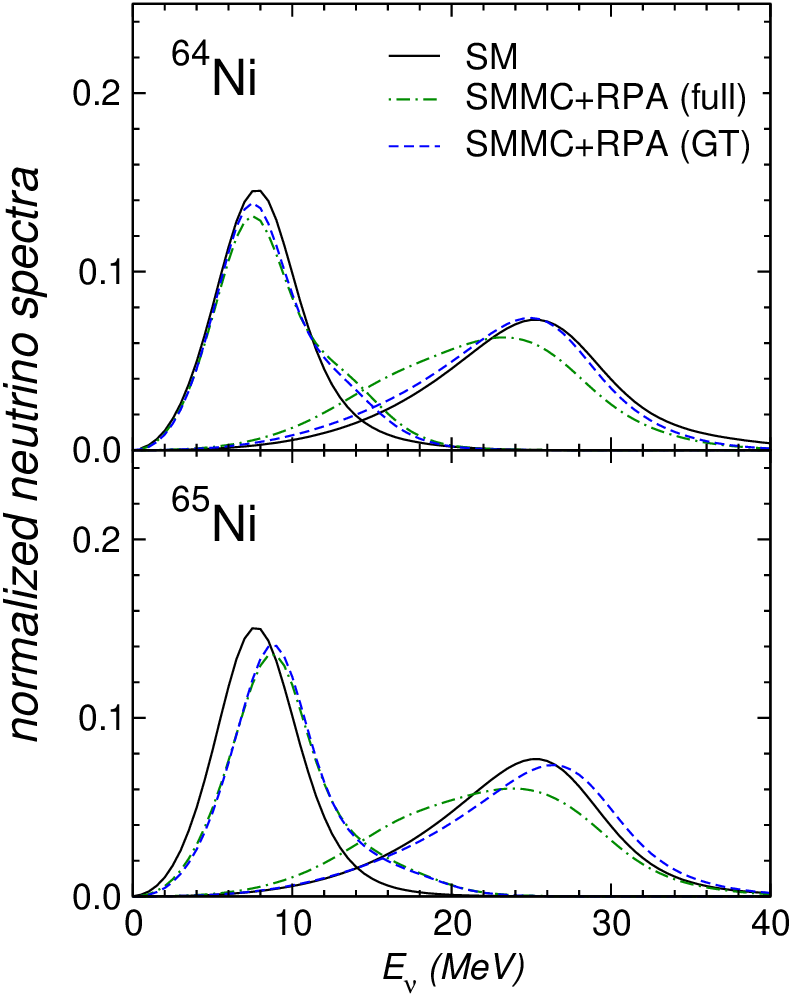}
\caption{\label{fig-NiIsotopes-Spectra}
(Color online) 
A comparison of the neutrino spectra emitted during electron capture on
$^{64,65}$Ni calculated from the diagonalization shell model (only allowed
contributions) and the hybrid SMMC+RPA model (both allowed and
forbidden contributions). Stellar conditions of two zones (10 and
15) of the $25\Msol$ trajectory (see Table \ref{tbl-Zones}) are
used. 
}
\end{center}
\end{figure}

The diagonalization shell model has been clearly established as the
method of choice to calculate electron capture rates, provided the
required model space is feasible with presently available computer
memory. The later restriction, however, prevents its use for many of
the heavier neutron-rich nuclei that are present in the collapsing
stellar core, forcing us to use the hybrid model as discussed
above. We have tested the validity and the accuracy of this hybrid
SMMC+RPA approach by comparing electron-capture rates, calculated
within this model, against diagonalization shell model results for two
nickel isotopes (Fig.~\ref{fig-NiIsotopes}). The calculations have
been performed for a set of stellar conditions (density, temperature,
$Y_e$ value) appropriate for the inner core of a collapsing
$25\Msol$ star (see Table~\ref{tbl-Zones}).  The shell model rates
are based only on the allowed Gamow-Teller (GT) transitions, while
the SMMC+RPA calculation also includes forbidden transitions. To
identify the effect of such forbidden transitions we have also
performed SMMC+RPA calculations considering only GT transitions.  As
can be seen in Fig.~\ref{fig-NiIsotopes}, the shell model and the
hybrid rates are quite similar for electron chemical potentials $\mue$
larger than about $10$~MeV when only the Gamow-Teller transitions
contribute to the electron-capture rate.  At lower $\mue$ values the
capture depends more sensitively on the detailed structure of the GT
strength distribution which is better described by the diagonalization
shell model than in the hybrid model. As explained in \cite{LMP03},
with increasing electron chemical potential the capture rate becomes
more and more dependent solely on the total GT strength rather than
on the details of its distribution. It is fortunate that
conditions that produce such low electron chemical potentials, for
which a detailed reproduction of the GT strength is required for a
reliable description of the rate, also result in relatively low mass
nuclei in NSE, for which diagonalization shell model calculations can
be performed and which are included in the LMP pool.  Forbidden
transitions start to contribute noticeably to the capture rates for
$\mue > 30$ MeV.  At such conditions corresponding to rather high
densities the nuclear composition is dominated by nuclei heavier than
included in the LMP pool and hence the neglect of higher multipole
transitions in the LMP rates is unimportant.

The reasonable reproduction of the shell-model Gamow-Teller
distribution by the SMMC+RPA approach is also reflected in the emitted
neutrino spectra, shown in Fig.~\ref{fig-NiIsotopes-Spectra}.  Once
the value of the electron chemical potential gets sufficiently large,
the emitted neutrino spectra obtained within the shell model and
SMMC+RPA approaches are quite similar.

\begin{figure}[tb]
\begin{center}
\includegraphics*[width=\figsize, angle=0]{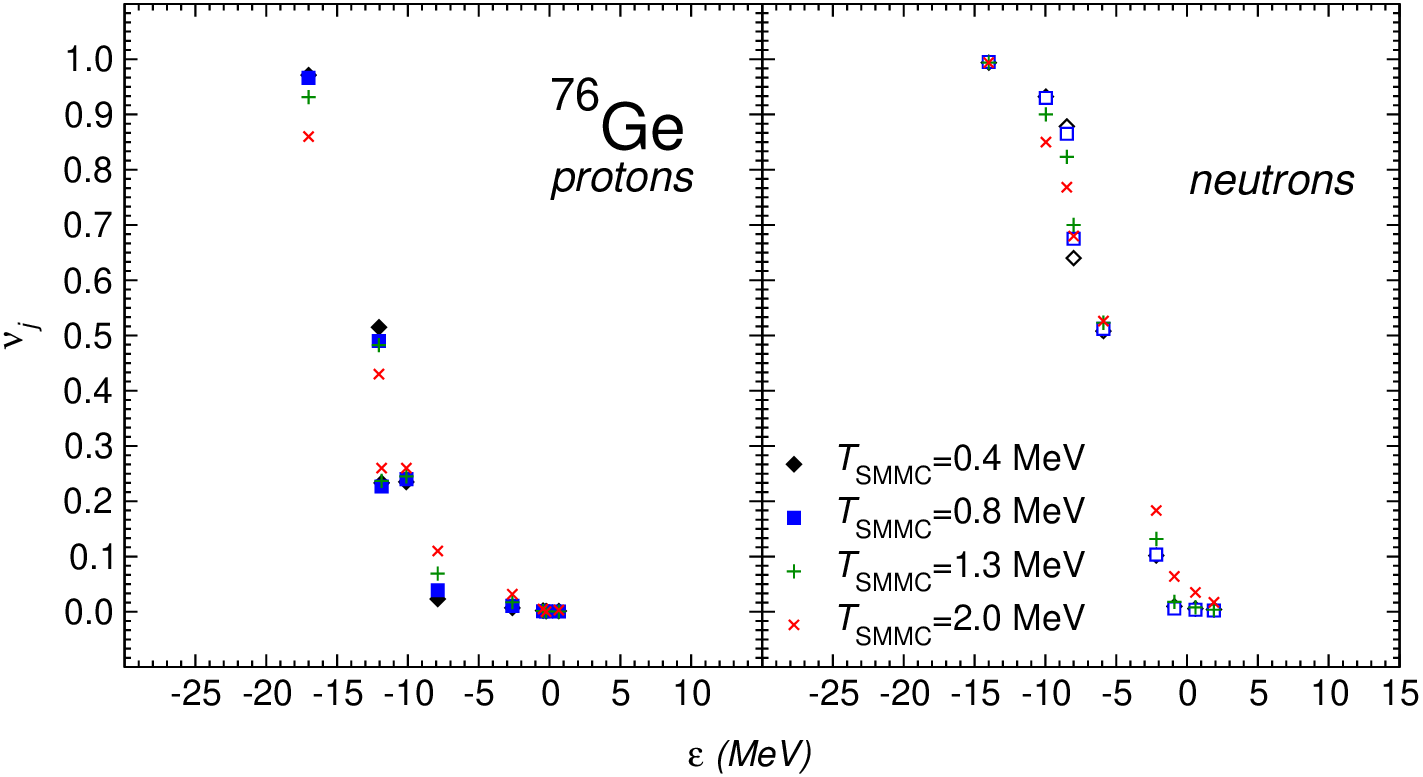}
\caption{(Color online) Single $j$-shell fractional occupation numbers
  in \updown{76}{32}Ge at different SMMC temperatures. Proton values
  are shown on the left, neutron values are shown on the right.  The
  fractional $j$-shell occupancy is calculated by $\nu_j=n_j/(2j+1)$,
  where $n_j$ is the number of particles in the shell. The
  single-particle energies are taken from the nucleus-adjusted
  Woods-Saxon potential.  }
\label{fig-occupation}
\end{center}
\end{figure}
\begin{figure}[tb]
\begin{center}
\includegraphics*[width=\figsize, angle=0]{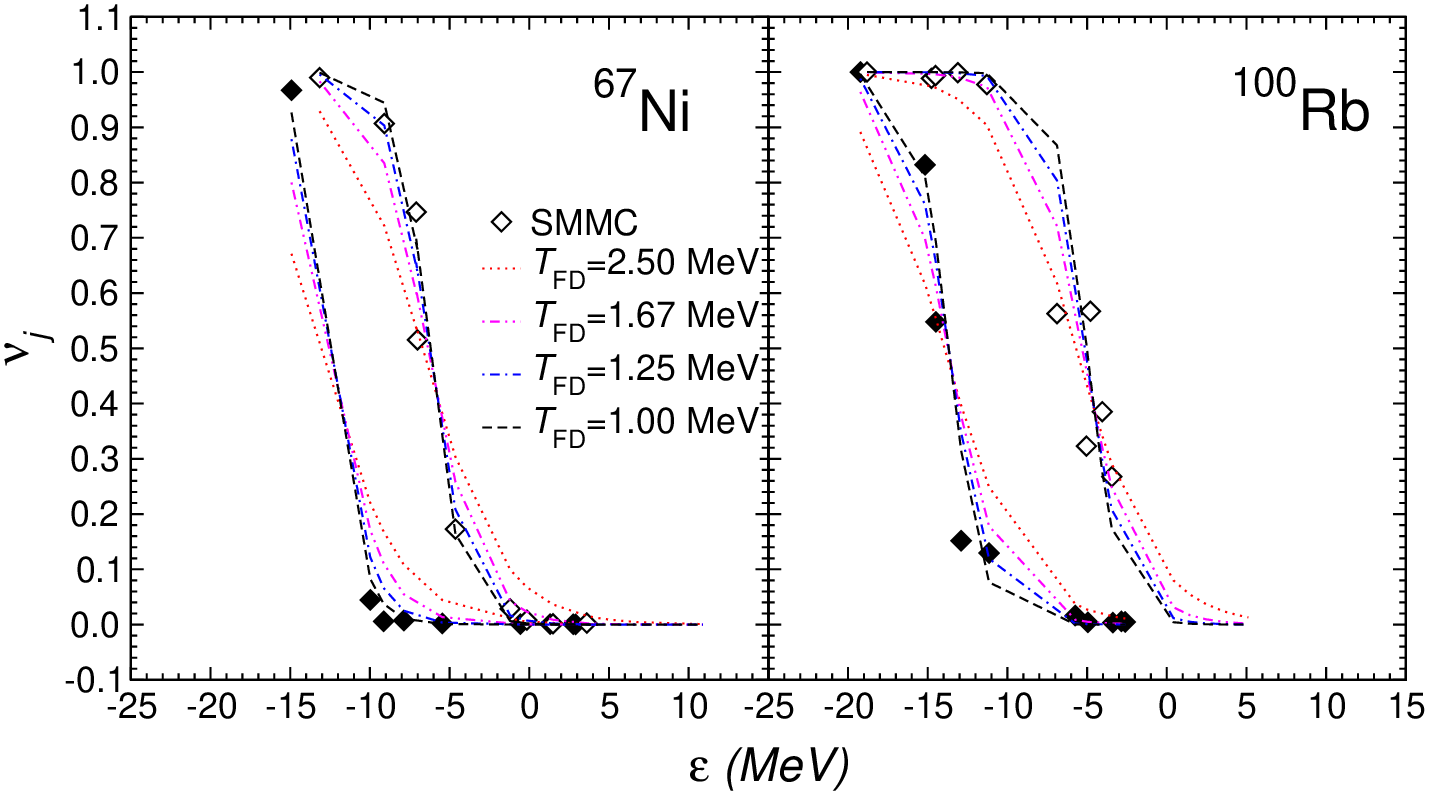}
\caption{(Color online)
Single $j$-shell fractional occupation numbers for
  \updown{67}{28}Ni and \updown{100}{\protect\phantom{1}37}Rb
  as a function of single-particle energies. 
  The SMMC results are shown by diamonds; 
  proton values are marked by full symbols, neutron values are marked
  by empty symbols. 
  The values for \updown{67}{28}Ni and \updown{100}{\protect\phantom{1}37}Rb
  were obtained at temperatures $T=0.50$ and $1.33$~MeV, respectively.
  The fractional occupations given by the
  Fermi-Dirac distribution for four values of $\TFD$ are shown by
  lines:
  $\TFD=2.50$~MeV (dotted), 
  $\TFD=1.67$~MeV (dash-double-dotted), 
  $\TFD=1.25$~MeV (dash-dotted),
  $\TFD=1.00$~MeV (dashed).
}\label{fig-Occs}
\end{center}
\end{figure}

A comment about the occupation numbers used in the SMMC+RPA
calculations is in order. In principle one should calculate these
numbers at all the temperatures needed to construct the stellar
capture rate table.  However, such a procedure is computationally untenable.  Fortunately it turns out that, within the range of
temperatures relevant to the phase of the collapse at which electron
captures are important ($T \approx 0.8$--$1.4$~MeV), the occupation
numbers do not vary too much. This is demonstrated in
Fig.~\ref{fig-occupation} which shows the proton and neutron
occupation numbers for $^{76}$Ge obtained by performing SMMC
calculations at different temperatures.  We conclude from this
comparison that it might suffice to calculate the SMMC occupation
numbers only at a single temperature which we chose as the value where
the respective nucleus has a large relative weight in the NSE
composition.

SMMC calculations are rather time-consuming. Thus calculations to
extend the SMMC pool beyond the $250$ nuclei considered here to
include the couple of thousand species which are present in the
supernova composition is prohibitive.  However, the observation that
the SMMC occupation numbers do not vary too much with temperature
motivated us to find a parameterized form for the SMMC occupation
numbers which could then readily be used in RPA calculations of
electron capture rates.  This goal is achieved by assuming a
Fermi-Dirac parameterization for the proton and neutron fractional
occupation numbers of various shells with energy $\varepsilon$: 
\beq
\nu_{\rm FD}(\varepsilon,\TFD)=\frac1{1+\exp
  \{(\varepsilon-\mu)/\TFD\}}.  
\eeq 

The chemical potentials are fixed by the total proton and neutron
numbers (here denoted by $N$)

\beq
N=\sum\limits_\varepsilon\, (2j_\varepsilon+1)\,\nu_{\rm
  FD}(\varepsilon,\TFD). \label{eq:fdchem} 
\eeq 

Here $j_\varepsilon$ denotes the total angular momentum of a shell
having the single-particle energy $\varepsilon$. These energies are
taken from a Woods-Saxon potential, the depth of which is adjusted
such as the neutron and proton chemical potentials resulting from
equation~(\ref{eq:fdchem}) equal the respective neutron and proton
separation energies, taken from either experiment or the compilation
of ref.~\cite{Moller95}.  The number of the considered single-$j$
shells depends on the numbers of nucleons and the value of the
temperature $\TFD$ and exceeds that used in the SMMC studies.  The
Fermi-Dirac distribution has one more undetermined parameter,
$\TFD$. We will vary it as a parameter below and fix its value by
attempting to reproduce the SMMC+RPA electron capture cross sections
for the $250$ nuclei present in the LMS pool.  In general we have
included at least $2$ major oscillator shells in our RPA calculations
outside a closed $N=Z=20$ core.

Fig.~\ref{fig-Occs} compares the SMMC occupation numbers for two
different nuclei with those obtained by an FD parameterization for
different values of the parameter $\TFD$.  The values $\TFD=1.0$ MeV
and $1.25$ MeV give a rather fair reproduction of the SMMC occupation
numbers for the two nuclei shown. (As we will see in the next chapter
the SMMC+RPA capture cross-sections are globally best reproduced by
the value $\TFD = 1.67$~MeV.)  It is important to note that the
occupation numbers reflect both correlation and thermal effects. Thus,
unlike for a system of non-interacting particles, the parameter $\TFD$
should not be interpreted as the temperature of the nucleus.
 
The stellar electron capture rate $\lambdaEC$ on a particular nucleus
is related to the electron capture cross-section $\sigmaEC$ by:
\begin{equation}
\lambdaEC = \frac{1}{\pi^2\hbar^3}\sum_{if}
\int_{\varepsilon_e^0}^{\infty}p_e^2
  \sigmaEC(\varepsilon_e,\varepsilon_i,\varepsilon_f)f(\varepsilon_e,\mu_e,T)d
  \varepsilon_e
\label{eq:ecrate}
\end{equation}
where $\varepsilon_e^0=\max(\Qif,m_e c^2)$, 
$p_e=(\varepsilon_e^2-m_e^2c^4)^{1/2}/c$ is the momentum of the
incoming electron with energy $\varepsilon_e$ and $m_e$ the electron
rest mass. Under the conditions present in the collapsing core of a
supernova, electrons obey a Fermi-Dirac distribution
$f(\varepsilon_e,\mu_e,T)=[1+\exp\{(\varepsilon_e-\mue)/{\kB}T\}]^{-1}$
with temperature 
$T$ and electron chemical potential $\mu_e$.
$\sigmaEC(\varepsilon_e,\varepsilon_i,\varepsilon_f)$ is the cross
section for capture of an electron with energy $\varepsilon_e$ from an
initial proton single particle state with energy $\varepsilon_i$ to a
neutron single particle state with energy $\varepsilon_f$. The cross
section is computed within the Random Phase Approximation as described
above. Due to energy conservation, the electron, proton and neutron
energies are related to the neutrino energy, $\varepsilon_\nu$, and
the $Q$-value for the 
capture reaction~\cite{Cooperstein84}:
\begin{equation}
  \label{eq:erpa}
  \Qif = \varepsilon_e - \varepsilon_\nu = \varepsilon^n_f -
  \varepsilon^p_i,
\end{equation}
\begin{equation}
  \label{eq:coop}
  \varepsilon^n_f -  \varepsilon^p_i = \varepsilon^*_{if} + \hat{\mu}
  + \Delta_{np}. 
\end{equation}
Here we take $\hat{\mu} = \mu_n - \mu_p$, the difference between
neutron and proton chemical potentials in the nucleus and $\Delta_{np}
= (M_n - M_p)c^2 = 1.293$~MeV, the neutron-proton mass difference. 

Equation~(\ref{eq:coop}) constitutes the definition of the quantity
$\varepsilon^*_{if}$. At zero temperature $\varepsilon^*_{if}$
corresponds to the excitation energy in the daughter nucleus. For the
transition from the initial ground state to the daughter ground state
the excitation energy must be zero. This fact is used for fixing
the value of $\hat{\mu}$:
\begin{equation}
  \label{eq:qmu}
  Q_{00} = (M_f - M_i)c^2 = \hat{\mu} + \Delta_{np},
\end{equation}
where $M_i$ and $M_f$ are the nuclear masses of the parent and
daughter nuclei in the electron capture reaction, and the values of
$\mu_p$ and $\mu_n$ are selected to equal proton and neutron
separation energies, $S_p(Z,N)$ and $S_n(Z-1,N+1)$.  At finite
temperature the value of $\varepsilon^*_{if}$ can be negative when the
nucleus is de-excited by capturing an electron.

The capture rate for individual nuclei is affected by screening
corrections that are implemented in our calculations as discussed in
the appendix. The screening corrections affect the electron capture in
two ways: they effectively increase the $Q$ value and lower the
electron chemical potential. As discussed in the next section and the
appendix, both effects lead to a reduction of the electron capture
rate in the medium as compared to the undisturbed case.

Calculation of the partial cross sections allows us to obtain the
spectrum of the emitted neutrinos. To have the spectrum normalized to
unity, we define it as follows:
\begin{equation}
  \nuprob =  
  \frac1{\lambdaEC}\, \frac{1}{\pi^2\hbar^3}\sum_{if} p_e^2
  \sigmaEC(\varepsilon_e,\varepsilon_i,\varepsilon_f) 
   f(\varepsilon_e,\mu_e,T),
   \label{eq-nuprob}
\end{equation}
with $\varepsilon_e = \varepsilon_f - \varepsilon_i + \varepsilon_\nu
$ and $p_e$ the corresponding electron
momentum. Equation~\refeq{eq-nuprob} defines the spectrum of emitted
neutrinos as produced by captures on a particular nucleus.

During stellar collapse, the matter is composed of individual
nuclei until densities of order $10^{13}$~\gcmic\ are reached.  While
presupernova studies (i.e.\ simulations which cover the late-stage
evolution of a massive star until the inner core has reached densities
up to $10^{9}$~\gcmic) require explicit consideration of the
extensive nuclear networks, temperatures in the subsequent supernova
collapse phase are high enough to bring reactions mediated by the
strong and electromagnetic force into equilibrium with their inverse
reactions and hence the nuclear composition can be described by
Nuclear Statistical Equilibrium (NSE). It is customary to split the
required electron capture rates for supernova simulations into rates
for protons $R_p=Y_p\lambda_p$, and for nuclei
$R_h=\sum_iY_i\lambda_i$ where the index $i$ runs over all isotopes
other than protons present in the stellar composition.  The required
abundances for protons $Y_p$ and for the various nuclear species $Y_i$
are derived from a Saha-like NSE distribution of the individual
isotopes where we included consideration of degenerate nucleons and
plasma corrections to the nuclear binding energy \cite{HT96}
using the screening
formula of Slattery {\em et al\/} \cite{Slattery82} with the
parameters from \cite{Ichimaru93} and the extension to the weak
screening regime as suggested by Yakovlev and Shalybkov 
\cite{Yakovlev.Shalybkov:1989} as
discussed in the appendix.  Compared to the LMSH tabulation used in
\cite{Langanke03,Hix03}, we have extended the pool of nuclei in the
NSE distribution to heavier nuclei (Ag to At).  Furthermore we have
used recently published partition functions which extend to 
temperatures in excess of $10$~GK \cite{Rauscher03}. The electron
chemical potential necessary for the calculation of the different
electron capture rates are determined using the {\sc helmholtz}
Equation of State by Timmes and Swesty \cite{TS00}.

In \cite{Langanke03,Hix03} the stellar electron capture
rate $\avgrate$ has been derived 
as the NSE-average over all nuclei for which individual rates
are available. Hence,
\beq
  \avgrate\,=\, \frac{\sum_i\,Y_i\lambda_i^{\rm ec}}{\sum_i\,Y_i}
\label{eq-avgrate}
\eeq 
Similarly, the NSE-averaged neutrino spectrum is 
obtained by calculating the ratio
\beq
  \avgdistr\  =\
  \frac1{\avgrate}\,
  \frac{\sum_i Y_i \nuprobi \lambda_i^{\rm ec}}{\sum_i Y_i}\
  =\
  \frac{\sum_i Y_i \nuprobi \lambda_i^{\rm ec}}{\sum_i Y_i\lambda_i^{\rm ec}}
\label{eq-avgdistr}
\eeq
This spectrum is normalized to unity, with the absolute neutrino
emission rate being $\avgrate\avgdistr$.

\begin{sidewaystable}[tbp]
\centering
\caption{\label{tbl-Zones}
Conditions for a central mass element in two progenitor stars at several
points in time during the collapse. $\bar A$ and $\bar Z$ is the average
mass and proton numbers as given by the NSE. $Y_p$ is the
abundance of protons. $Y_h$ is the abundance of heavy nuclei defined
as $Y_h=(\sum Y_i)-Y_p-Y_n-Y_\alpha$, i.e.\ everything except for
protons, neutrons and alpha particles. 
$Z_h$ and $A_h$ are the average mass and charge of the heavy nuclei.
The last two columns list values of the unscreened and screened 
electron capture rates ($\avgrateNS$ and $\avgrate$, respectively).
}
\resizebox{\textwidth}{!}{
\begin{tabular}{cccddddddddddcc}
\hline\hline
\MC{1}{c}{idx} & \MC{1}{c}{$\rho$} & \MC{1}{c}{$Y_e$} & \MC{1}{c}{$T$} & \MC{1}{c}{$k_BT$} & \MC{1}{c}{$\mu_e$} & \MC{1}{c}{${\widebar A}$} & \MC{1}{c}{${\widebar Z}$}& \MC{1}{c}{${\widebar A}_h$} & \MC{1}{c}{${\widebar Z}_h$} & \MC{1}{c}{$Y_p$} & \MC{1}{c}{$Y_h$} & \MC{1}{c}{$Y_p/Y_h$} & \MC{1}{c}{$\avgrateNS$} & \MC{1}{c}{$\avgrate$} \\
  & \MC{1}{c}{(\gcmic)} & \MC{1}{c}{} & \MC{1}{c}{(GK)} & \MC{1}{c}{(MeV)} & \MC{1}{c}{(MeV)} & & & & & \MC{1}{c}{(mol/g)} & \MC{1}{c}{(mol/g)} & & \MC{1}{c}{(\secinv)} & \MC{1}{c}{(\secinv)}
\\ \hline
\MC{10}{c}{\it $15\Msol$ progenitor\/}  & \MC{1}{r}{$\times10^{-5}$} & \MC{1}{r}{$\times10^{-2}$} & \MC{1}{r}{$\times10^{-3}$} \\
  1 &  $8.56\times 10^{ 9}$ &  0.434 &  7.49 &  0.65 &   7.83  &  58.0 &  25.2 &  60.2 &  26.1 &  0.230 & 1.66 & 0.139 &  $1.94\times 10^{ 0}$ & $1.38\times 10^{ 0}$ \\
  2 &  $9.34\times 10^{ 9}$ &  0.434 &  7.73 &  0.67 &   8.06  &  57.4 &  24.9 &  60.3 &  26.2 &  0.349 & 1.65 & 0.211 &  $2.62\times 10^{ 0}$ & $1.87\times 10^{ 0}$ \\
  3 &  $1.34\times 10^{10}$ &  0.431 &  8.78 &  0.76 &   9.04  &  52.1 &  22.5 &  61.2 &  26.4 &  1.61  & 1.62 & 0.991 &  $8.15\times 10^{ 0}$ &  $5.83\times 10^{ 0}$ \\
  4 &  $2.42\times 10^{10}$ &  0.422 & 10.02 &  0.88 &  10.97  &  40.8 &  17.2 &  64.6 &  27.4 &  5.24  & 1.52 & 3.44  &  $4.10\times 10^{ 1}$ &  $2.96\times 10^{ 1}$ \\
  5 &  $4.71\times 10^{10}$ &  0.403 & 11.39 &  0.98 &  13.53  &  30.1 &  12.1 &  73.2 &  30.0 &  4.22  & 1.33 & 3.17  &  $1.47\times 10^{ 2}$ &  $1.05\times 10^{ 2}$ \\
  6 &  $6.95\times 10^{10}$ &  0.391 & 11.91 &  1.03 &  15.27  &  24.1 &   9.4 &  77.1 &  30.9 &  2.50  & 1.26 & 2.00  &  $2.14\times 10^{ 2}$ &  $1.50\times 10^{ 2}$ \\
  7 &  $1.12\times 10^{11}$ &  0.377 & 12.64 &  1.09 &  17.76  &  17.9 &   6.7 &  79.1 &  31.1 &  1.53  & 1.21 & 1.27  &  $3.31\times 10^{ 2}$ &  $2.24\times 10^{ 2}$ \\
  8 &  $1.83\times 10^{11}$ &  0.365 & 13.57 &  1.17 &  20.69  &  14.0 &   5.1 &  79.9 &  31.0 &  1.33  & 1.17 & 1.13  &  $6.96\times 10^{ 2}$ &  $4.50\times 10^{ 2}$ \\
  9 &  $2.33\times 10^{11}$ &  0.359 & 14.11 &  1.22 &  22.35  &  12.4 &   4.5 &  80.3 &  31.0 &  1.32  & 1.16 & 1.14  &  $1.10\times 10^{ 3}$ &  $6.97\times 10^{ 2}$ \\
 10 &  $3.76\times 10^{11}$ &  0.344 & 15.26 &  1.32 &  25.84  &   9.6 &   3.3 &  81.7 &  30.9 &  1.17  & 1.11 & 1.05  &  $2.68\times 10^{ 3}$ &  $1.64\times 10^{ 3}$ \\
 11 &  $5.89\times 10^{11}$ &  0.319 & 16.22 &  1.40 &  29.33  &   7.1 &   2.3 &  85.7 &  31.5 &  0.646 & 1.01 & 0.637 &  $5.70\times 10^{ 3}$ &  $3.36\times 10^{ 3}$ \\
 12 &  $9.33\times 10^{11}$ &  0.295 & 17.51 &  1.51 &  33.31  &   5.5 &   1.6 &  93.0 &  33.2 &  0.471 & 0.888 & 0.530 & $1.37\times 10^{ 4}$ &  $7.76\times 10^{ 3}$ \\
 13 &  $1.53\times 10^{12}$ &  0.284 & 19.82 &  1.71 &  38.79  &   5.0 &   1.4 &  98.1 &  34.5 &  0.884 & 0.823 & 1.07  & $4.67\times 10^{ 4}$ &  $2.60\times 10^{ 4}$ \\
 14 &  $1.95\times 10^{12}$ &  0.280 & 20.99 &  1.81 &  41.82  &   4.8 &   1.4 & 100.2 &  34.9 &  1.09  & 0.799 & 1.36  &  $8.30\times 10^{ 4}$ &  $4.60\times 10^{ 4}$ \\
 15 &  $2.47\times 10^{12}$ &  0.275 & 22.37 &  1.93 &  45.04  &   4.7 &   1.3 & 101.5 &  35.1 &  1.40  & 0.782 & 1.79  &  $1.46\times 10^{ 5}$ &  $8.06\times 10^{ 4}$ \\
 16 &  $3.17\times 10^{12}$ &  0.272 & 24.08 &  2.08 &  48.76  &   4.6 &   1.3 & 101.8 &  34.9 &  2.04  & 0.777 & 2.63  &  $2.72\times 10^{ 5}$ &  $1.51\times 10^{ 5}$ \\
 17 &  $4.04\times 10^{12}$ &  0.269 & 25.54 &  2.20 &  52.72  &   4.6 &   1.2 & 103.9 &  35.3 &  2.32  & 0.762 & 3.05  &  $4.91\times 10^{ 5}$ &  $2.70\times 10^{ 5}$ \\
 18 &  $5.14\times 10^{12}$ &  0.263 & 27.06 &  2.33 &  56.70  &   4.5 &   1.2 & 106.8 &  35.8 &  2.39  & 0.736 & 3.25  &  $8.23\times 10^{ 5}$ &  $4.47\times 10^{ 5}$ \\
 19 &  $6.59\times 10^{12}$ &  0.259 & 29.43 &  2.54 &  61.27  &   4.5 &   1.2 & 106.5 &  35.2 &  3.32  & 0.735 & 4.52  &  $1.47\times 10^{ 6}$ &  $7.92\times 10^{ 5}$ \\
 20 &  $8.51\times 10^{12}$ &  0.261 & 32.12 &  2.77 &  66.88  &   4.7 &   1.2 & 105.6 &  34.7 &  4.98  & 0.750 & 6.63  &  $2.99\times 10^{ 6}$ &  $1.58\times 10^{ 6}$ \\
 21 &  $1.07\times 10^{13}$ &  0.260 & 33.76 &  2.91 &  72.14  &   4.8 &   1.2 & 111.3 &  36.1 &  4.61  & 0.719 & 6.41  &  $5.18\times 10^{ 6}$ &  $2.67\times 10^{ 6}$ \\
\hline\hline
\end{tabular}
}
\end{sidewaystable}

\addtocounter{table}{-1}
\begin{sidewaystable}[tbp]
\begin{flushright}
{\em Continuation of Table~\ref{tbl-Zones}}
\end{flushright}
\centering
\resizebox{\textwidth}{!}{
\begin{tabular}{cccddddddddddcc}
\hline\hline
\MC{1}{c}{idx} & \MC{1}{c}{$\rho$} & \MC{1}{c}{$Y_e$} & \MC{1}{c}{$T$} & \MC{1}{c}{$k_BT$} & \MC{1}{c}{$\mu_e$} & \MC{1}{c}{${\widebar A}$} & \MC{1}{c}{${\widebar Z}$}& \MC{1}{c}{${\widebar A}_h$} & \MC{1}{c}{${\widebar Z}_h$} & \MC{1}{c}{$Y_p$} & \MC{1}{c}{$Y_h$} & \MC{1}{c}{$Y_p/Y_h$} & \MC{1}{c}{$\avgrateNS$} & \MC{1}{c}{$\avgrate$} \\
  & \MC{1}{c}{(\gcmic)} & \MC{1}{c}{} & \MC{1}{c}{(GK)} & \MC{1}{c}{(MeV)} & \MC{1}{c}{(MeV)} & & & & & \MC{1}{c}{(mol/g)} & \MC{1}{c}{(mol/g)} & & \MC{1}{c}{(\secinv)} & \MC{1}{c}{(\secinv)}
\\ \hline
\MC{10}{c}{\it $25\Msol$ progenitor\/} & \MC{1}{r}{$\times10^{-4}$} & \MC{1}{r}{$\times10^{-2}$} & \MC{1}{r}{$\times10^{-2}$} \\
  1 &  $2.22\times 10^{ 9}$ &  0.447 &  7.86 &  0.68 &   4.87  &  44.8 &  20.0 &  56.4 &  25.1 &  0.788 & 1.74 & 0.451 &  $8.99\times 10^{-2}$ &  $6.65\times 10^{-2}$ \\
  2 &  $2.97\times 10^{ 9}$ &  0.445 &  8.44 &  0.73 &   5.38  &  39.3 &  17.5 &  56.5 &  25.1 &  1.46  & 1.72 & 0.845 &  $2.04\times 10^{-1}$ &  $1.51\times 10^{-1}$ \\
  3 &  $5.59\times 10^{ 9}$ &  0.435 &  9.83 &  0.85 &   6.61  &  25.4 &  11.0 &  57.4 &  25.1 &  3.09  & 1.64 & 1.88  &  $9.77\times 10^{-1}$ &  $7.32\times 10^{-1}$ \\
  4 &  $6.74\times 10^{ 9}$ &  0.431 & 10.22 &  0.88 &   7.03  &  22.3 &   9.6 &  57.9 &  25.1 &  3.56  & 1.61 & 2.21  &  $1.58\times 10^{ 0}$ &  $1.19\times 10^{ 0}$ \\
  5 &  $9.79\times 10^{ 9}$ &  0.425 & 10.81 &  0.93 &   7.96  &  19.4 &   8.2 &  58.9 &  25.4 &  3.92  & 1.57 & 2.50  &  $4.04\times 10^{ 0}$ &  $3.02\times 10^{ 0}$ \\
  6 &  $1.46\times 10^{10}$ &  0.418 & 11.44 &  0.99 &   9.09  &  17.1 &   7.1 &  60.5 &  25.8 &  4.09  & 1.52 & 2.69  &  $1.08\times 10^{ 1}$ &  $7.99\times 10^{ 0}$ \\
  7 &  $2.49\times 10^{10}$ &  0.407 & 12.10 &  1.04 &  10.84  &  16.0 &   6.5 &  64.2 &  26.9 &  3.15  & 1.44 & 2.19  &  $3.73\times 10^{ 1}$ &  $2.74\times 10^{ 1}$ \\
  8 &  $4.26\times 10^{10}$ &  0.395 & 12.72 &  1.10 &  12.92  &  15.3 &   6.0 &  69.1 &  28.4 &  2.04  & 1.35 & 1.51  &  $1.09\times 10^{ 2}$ &  $7.92\times 10^{ 1}$ \\
  9 &  $6.91\times 10^{10}$ &  0.384 & 13.35 &  1.15 &  15.10  &  14.0 &   5.4 &  73.3 &  29.6 &  1.36  & 1.27 & 1.07  &  $2.41\times 10^{ 2}$ &  $1.71\times 10^{ 2}$ \\
 10 &  $1.13\times 10^{11}$ &  0.372 & 14.11 &  1.22 &  17.68  &  12.2 &   4.5 &  76.3 &  30.4 &  0.934 & 1.21 & 0.772 &  $4.85\times 10^{ 2}$ &  $3.35\times 10^{ 2}$ \\
 11 &  $1.88\times 10^{11}$ &  0.361 & 15.11 &  1.30 &  20.75  &  10.3 &   3.7 &  77.9 &  30.6 &  0.752 & 1.17 & 0.644 &  $1.07\times 10^{ 3}$ &  $7.10\times 10^{ 2}$ \\
 12 &  $3.75\times 10^{11}$ &  0.339 & 16.76 &  1.44 &  25.66  &   7.8 &   2.6 &  79.9 &  30.6 &  0.532 & 1.10 & 0.483 &  $3.42\times 10^{ 3}$ &  $2.13\times 10^{ 3}$ \\
 13 &  $5.92\times 10^{11}$ &  0.315 & 17.79 &  1.53 &  29.17  &   6.0 &   1.9 &  83.1 &  30.9 &  0.291 & 1.01 & 0.287 &  $6.81\times 10^{ 3}$ &  $4.08\times 10^{ 3}$ \\
 14 &  $9.41\times 10^{11}$ &  0.291 & 19.32 &  1.67 &  33.23  &   4.9 &   1.4 &  87.8 &  31.8 &  0.234 & 0.915 & 0.256 &  $1.64\times 10^{ 4}$ &  $9.54\times 10^{ 3}$ \\
 15 &  $1.54\times 10^{12}$ &  0.281 & 21.84 &  1.88 &  38.68  &   4.5 &   1.3 &  90.1 &  32.1 &  0.376 & 0.871 & 0.432 &  $5.41\times 10^{ 4}$ &  $3.10\times 10^{ 4}$ \\
 16 &  $1.97\times 10^{12}$ &  0.275 & 23.21 &  2.00 &  41.74  &   4.3 &   1.2 &  90.7 &  32.0 &  0.446 & 0.858 & 0.520 &  $9.53\times 10^{ 4}$ &  $5.43\times 10^{ 4}$ \\
 17 &  $2.53\times 10^{12}$ &  0.271 & 24.90 &  2.15 &  45.09  &   4.2 &   1.1 &  89.6 &  31.4 &  0.587 & 0.861 & 0.682 &  $1.70\times 10^{ 5}$ &  $9.70\times 10^{ 4}$ \\
 18 &  $3.25\times 10^{12}$ &  0.269 & 26.78 &  2.31 &  48.90  &   4.2 &   1.1 &  87.9 &  30.5 &  0.777 & 0.876 & 0.887 &  $3.18\times 10^{ 5}$ &  $1.81\times 10^{ 5}$ \\
 19 &  $4.14\times 10^{12}$ &  0.265 & 28.43 &  2.45 &  52.77  &   4.2 &   1.1 &  88.0 &  30.2 &  0.832 & 0.874 & 0.952 &  $5.47\times 10^{ 5}$ &  $3.09\times 10^{ 5}$ \\
 20 &  $5.28\times 10^{12}$ &  0.259 & 30.43 &  2.62 &  56.84  &   4.1 &   1.1 &  87.0 &  29.5 &  0.916 & 0.877 & 1.04  &  $9.20\times 10^{ 5}$ &  $5.14\times 10^{ 5}$ \\
 21 &  $6.79\times 10^{12}$ &  0.257 & 33.24 &  2.86 &  61.65  &   4.1&   1.1  &  82.9 &  27.8 &  1.26 & 0.922 & 1.37  &  $1.71\times 10^{ 6}$ &  $9.47\times 10^{ 5}$ \\
 22 &  $8.75\times 10^{12}$ &  0.258 & 35.99 &  3.10 &  67.18  &   4.3 &   1.1 &  81.3 &  27.0 &  1.58 & 0.952 & 1.65  &  $3.33\times 10^{ 6}$ &  $1.82\times 10^{ 6}$ \\
 23 &  $1.11\times 10^{13}$ &  0.256 & 37.99 &  3.27 &  72.40  &   4.4 &   1.1 &  84.8 &  27.7 &  1.42 & 0.919 & 1.55  &  $5.59\times 10^{ 6}$ &  $2.98\times 10^{ 6}$ \\
\hline\hline
\end{tabular}
}
\end{sidewaystable}

However, there is a potential problem with the average rate as defined
above, when applied to the pool of nuclei used in
\cite{Langanke03,Hix03}, as this pool was limited to nuclei with
mass numbers $A<112$, sampling only a rather small fraction of the
total NSE abundance in the later phase of the collapse 
(changes in nuclear composition during the collapse are
illustrated in \cite{Hix03JPG}).
While this averaging of the rate prevents the total rate of electron 
capture from unphysically dropping to zero as the fraction of nuclei with 
calculated rates declines, the average rate becomes increasingly dominated 
by the species at the neutron-rich edge of the calculated pools, making the 
averaged rates of~\cite{Langanke03,Hix03} uncertain. It is the aim of the 
present work to overcome this shortcoming by appropriately enlarging the pool 
of nuclei from which the averages \refeq{eq-avgrate} and
\refeq{eq-avgdistr} are being derived. The present pool contains
nearly $2700$ nuclei (in contrast to about $280$ considered in
\cite{Langanke03,Hix03}).  For nuclei with mass numbers $A<65$ our
pool consists of the LMP shell model rates
\cite{Langanke.Martinez-Pinedo:2001} (i.e.\ the $pf$-shell nuclei)
supplemented by the FFN rates for the lighter $sd$-shell nuclei.  We have also
extended the pool of nuclei for which rates are based on the hybrid
SMMC+RPA approach to include nuclei up to $A=120$ (SMMC
pool).  The pool is completed by more than $2500$ nuclei for which we
have derived rates based on RPA calculations with the occupation
numbers approximated by the parameterized FD distribution as discussed
above (FD pool). As some of the nuclei calculated within this approach 
are overlapping with the nuclei from other pools, we do not use the 
FD+RPA results if the rates and spectra are already provided by other 
pools, unless specifically indicated.

In the next section we present electron capture rates and the emitted
neutrino spectra for conditions during the core collapse of two
progenitor stars with different masses.  The first set of conditions
follows a central mass element (at an enclosed mass of $0.05$ solar
masses) during the collapse of a $15\Msol$ progenitor star derived
from a simulation using electron-capture rates from the LMSH
tabulation \cite{Langanke03,Hix03} and general relativity.  The
second set is based on a similar simulation, but for a $25\Msol$
progenitor star. The stellar conditions are defined in Table~\ref{tbl-Zones}.

\section{Results and discussion}
\label{sect-Results}

\begin{figure}[tbp]
\begin{center}
\includegraphics*[width=\figsize, angle=0]{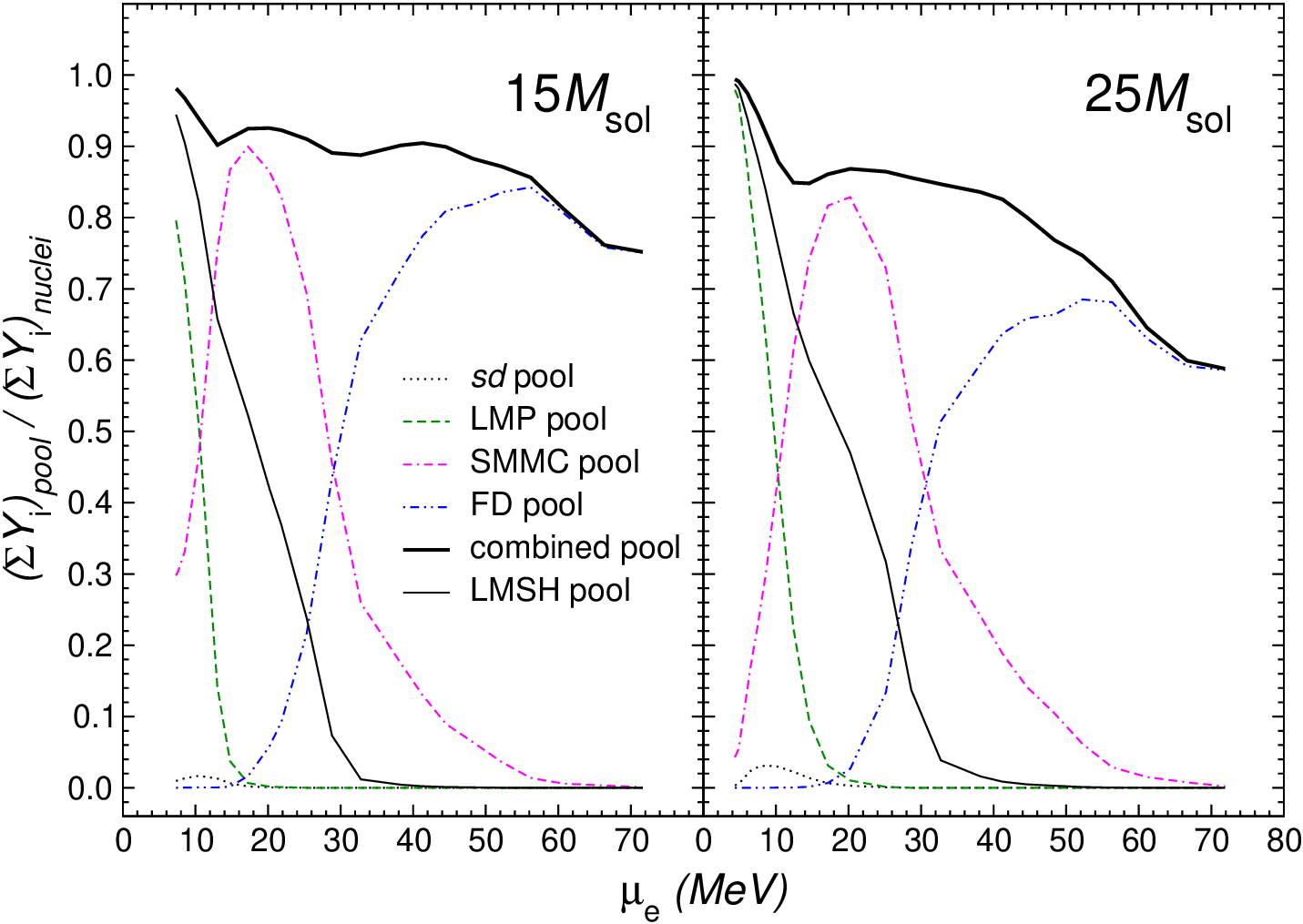}
\caption{ \label{fig-Coverage} (Color online)
  Fraction of nuclei covered by the various pools of nuclei as defined in the text. The fractions have been calculated for 
  the two stellar trajectories given 
  in  Table~\ref{tbl-Zones}.
  The pools are $sd$  (dotted line), LMP (dashed line),
  SMMC+RPA (double-dash-dotted line), and FD+RPA (dash-double-dotted line).
Solid lines show the summed pool coverage. 
Thick solid lines show present pool coverage, and thin solid lines
show coverage by the LMSH pool.
$(\sum_iY_i)_{\rm nuclei}$ is calculated by
summing over all nuclei except protons, neutrons and $\alpha$ particles.}
\end{center}
\end{figure}

\begin{figure}[tbp]
\begin{center}
\includegraphics*[width=\figsizeBig, angle=0]{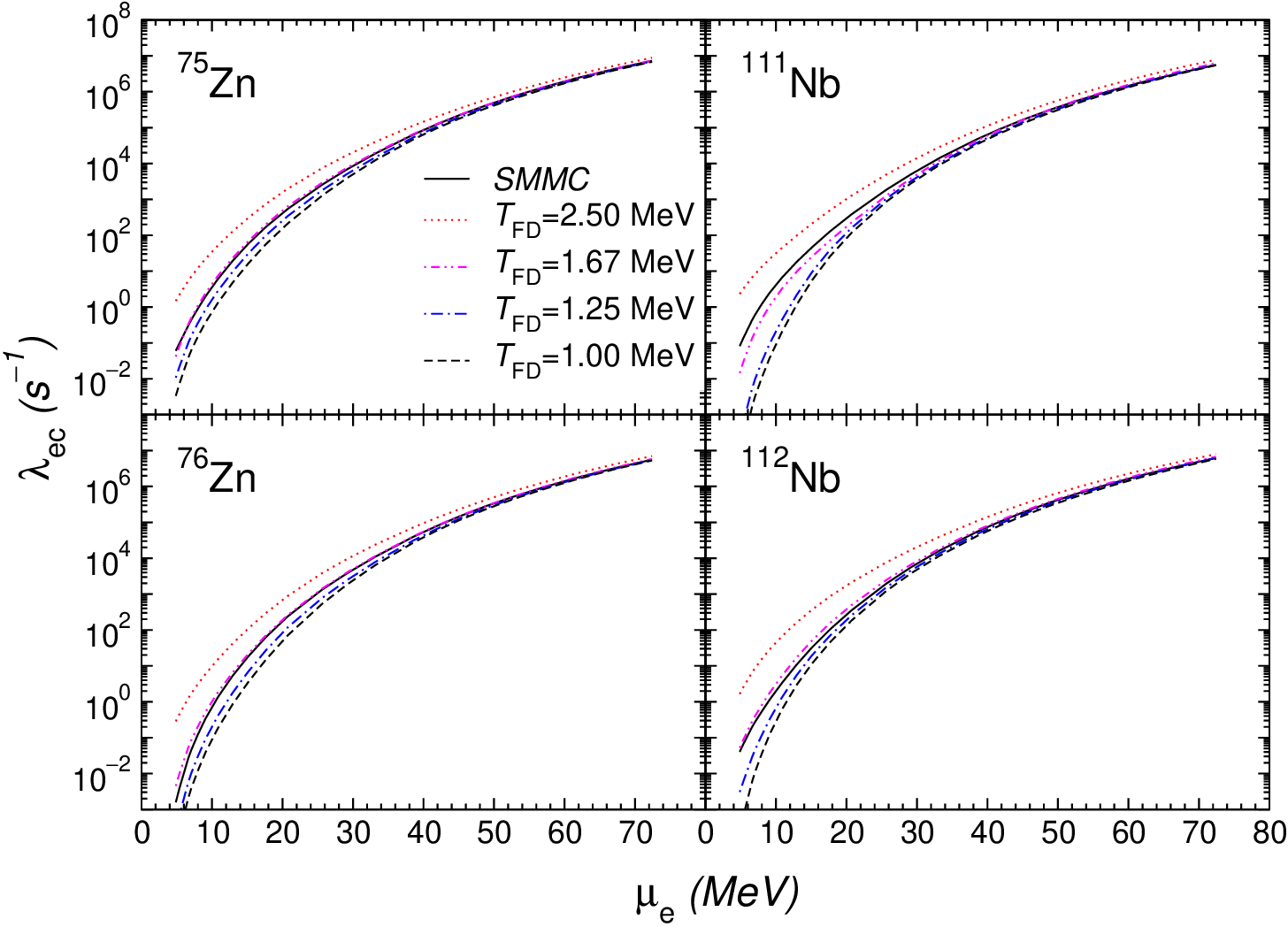}
\caption{\label{fig-IndRates} (Color online) Electron capture
  rates for four
  nuclei from the SMMC pool: $^{75,76}$Zn, $^{111,112}$Nb. 
  The SMMC+RPA rates 
  are shown by solid lines. The FD+RPA rates have been 
  calculated for $\TFD=2.50$ MeV (dotted line), 
  $\TFD=1.67$ MeV (dash-double-dotted line), 
  $\TFD=1.25$ MeV (dash-dotted line), and
  $\TFD=1.00$ MeV (dashed line).
  The stellar conditions are those of $25\Msol$
  progenitor star (Table~\ref{tbl-Zones}).
} 
\end{center}
\end{figure}

\begin{figure}[tbp]
\begin{center}
\includegraphics*[width=\figsizeBig, angle=0]{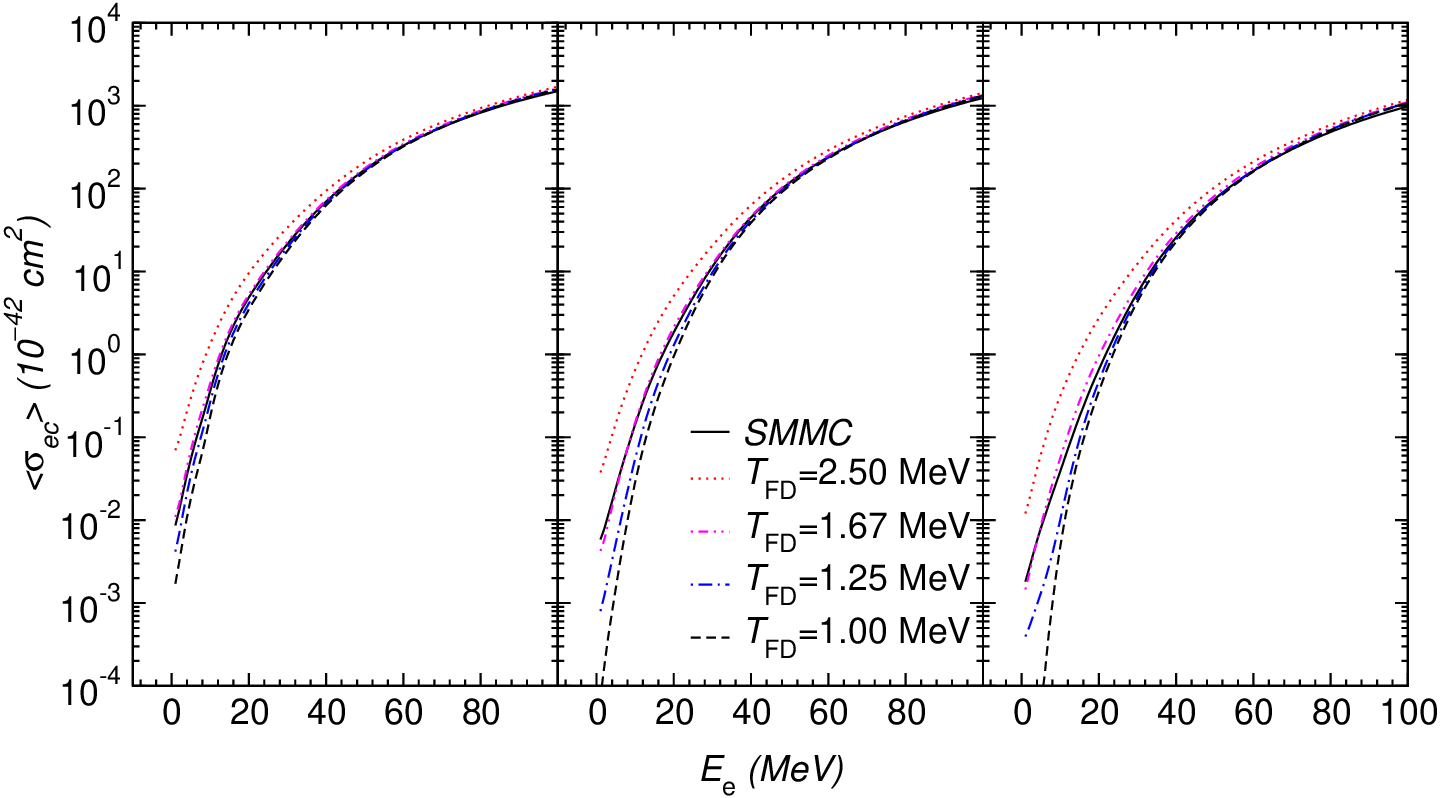}
\caption{\label{fig-AvgXSect-diffBeta} (Color online) NSE-averaged
  electron-capture cross-sections as calculated
  by the SMMC+RPA and FD+RPA approaches for the $250$ nuclei of the SMMC+RPA
  pool at three stellar conditions of the $25\Msol$ trajectory:
  the snapshot number 4 is used in the left panel, the number 10 in
  the center panel, and the number 13 in the right panel.  
  The notation is the same as in Fig.\ \ref{fig-IndRates}.
}
\end{center}
\begin{center}
\includegraphics*[width=\figsize, angle=0]{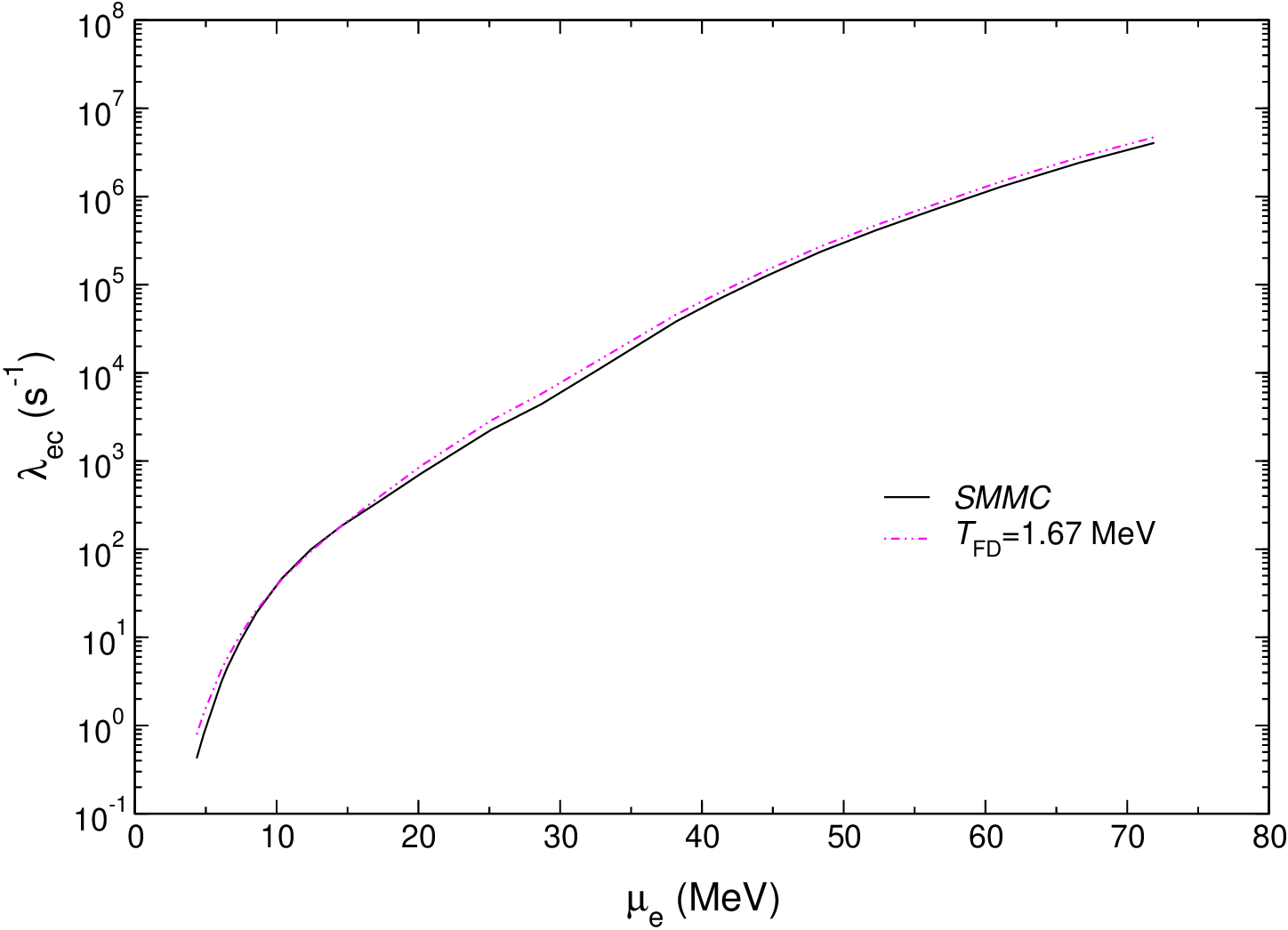}
\caption{\label{fig-AvgRate-diffBeta} (Color online)
  NSE-averaged electron-capture rates as calculated
  by the SMMC+RPA and FD+RPA approaches for the $250$ nuclei of the
  SMMC+RPA pool. The stellar conditions are those of $25\Msol$
  progenitor star (Table~\ref{tbl-Zones}).
  The solid line shows  the SMMC+RPA rate. 
  The dashed line gives the FD+RPA rates 
  calculated with $\TFD=1.67$ MeV. 
  }
\end{center}
\end{figure}

\begin{figure}[tbp]
\begin{center}
\includegraphics*[width=\figsizeBig, angle=0]{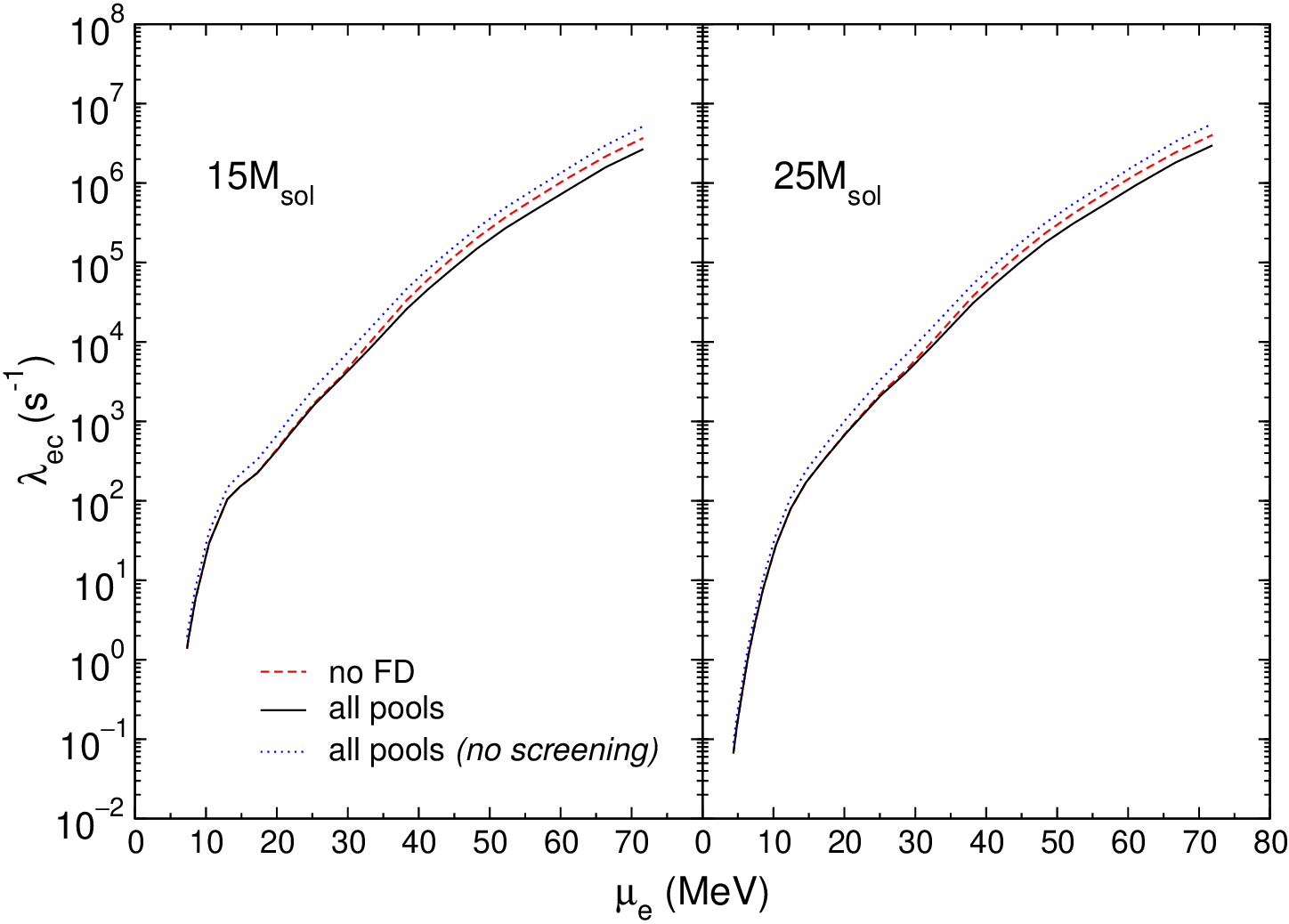}
\caption{ \label{fig-AvgRate} (Color online)
  Pool-averaged electron-capture rates calculated along the
  stellar trajectories for the $15\Msol$ and
  $25\Msol$ progenitor stars. The rates based on the sum of all pools 
  of nuclei
  are shown by solid lines. The dashed lines show the average rate
  when the FD+RPA pool is omitted. The dotted lines show the average 
  rate for the sum of all pools when the screening 
  effects to the rates are neglected.
}
\end{center}
\end{figure}
\begin{figure}[tbp]
\begin{center}
\includegraphics*[width=\figsize, angle=0]{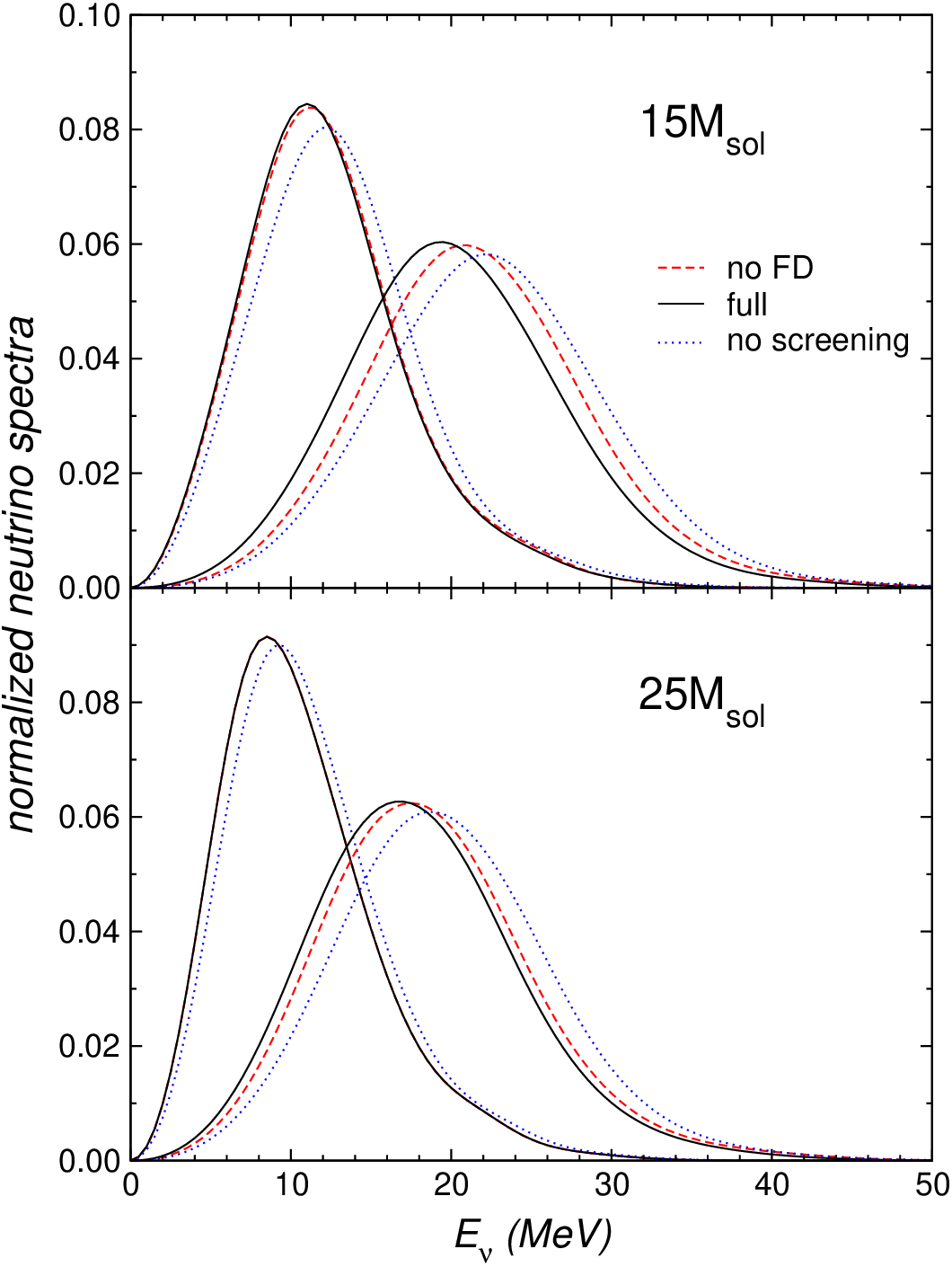}
\caption{ \label{fig-AvgSpectra} (Color online) Pool-averaged emitted
  neutrino spectra for the $15\Msol$ and $25\Msol$ trajectories.  The
  line legend is the same as in Fig.~\ref{fig-AvgRate}.  Two stellar
  conditions are used in each case corresponding to snapshot numbers
  10 and 15 of the respective trajectory. For snapshot number 10 in
  the lower panel the curves ``no FD'' and ``full'' coincide.}
\end{center}
\end{figure}

As mentioned above, the pool of nuclei adopted to derive the LMSH electron 
capture tabulation used in \cite{Langanke03,Hix03} was limited to isotopes
with $A<112$ and turns out to cover only a small fraction of the total
NSE abundance in the latest stages of the collapse for two reasons.
First, heavier nuclei appear in the nuclear composition with
noticeable abundances. Secondly, the inclusion of electron capture 
on nuclei makes the matter more neutron-rich than anticipated in prior
studies which considered only capture on protons and neglected the
dominant electron reducing weak process. As the pool of nuclei 
used in \cite{Langanke03} has been
constructed based on trajectories from prior collapse simulations which solely
considered capture on protons, the pool was missing relevant 
neutron-rich isotopes. 

The limitation of the nuclear pool used in \cite{Langanke03} 
(LMSH pool, i.e.\ the combined $sd$, LMP and LMS pools)
is shown in Fig.~\ref{fig-Coverage}.
This sample is complete for the early stage of the collapse ($\simlt 10^{10}$ \gcmic) where the pool includes more than $90\%$ of the nuclei present in 
the supernova medium (indices 1-3 of the $15\Msol$ and 1-4 of the $25\Msol$
trajectories in Table~\ref{tbl-Zones}).  However, the coverage drops quickly 
as the collapse progresses. The pool covers more than $50\%$ of the 
total NSE abundance for densities $\simlt 10^{11}$ \gcmic (indices 7 and 10 respectively). At the conditions around neutrino trapping (indices 10 and 12, respectively, of the $15\Msol$ and $25\Msol$ collapse trajectories) only $20\%$
of the total NSE abundance is sampled by the LMSH pool.  This
drastic underrepresentation is improved slightly by including the
additional $170$ nuclei for which we have calculated capture rates
using the SMMC+RPA approach, pushing the limits at which the $20\%$
coverage of the total NSE abundance is reached to indices 11 and 15
for the $15\Msol$ and $25\Msol$ trajectories, respectively.

Obviously improving the coverage requires an appropriate enlargement of 
the nuclear pool from which the capture rates are derived.
This is in particular motivated by the fact that 
the omission of the neutron-rich
nuclei from the pool in \cite{Langanke03}
could result in a systematic overestimate of the
rates as the capture process gets increasingly hindered
with growing neutron excess due to Pauli blocking.
However, covering the relevant nuclei by SMMC+RPA calculations is 
too computationally demanding. Thus we have performed evaluations of
the capture rates for more than $2200$ nuclei using occupation
numbers derived from a FD parameterization. These nuclei consider the
isotope chains for charge numbers $Z=28$--$70$. When these nuclei are
added to the pool, Fig.~\ref{fig-Coverage} shows that at least $60\%$
of the total NSE abundance is considered during the entire collapse
evolution until densities of order $10^{13}$ \gcmic\ are reached and
the description of the nuclear composition by individual nuclei is
inadequate.  We stress that at the beginning of the core collapse, the
center of a progenitor star is populated by iron group nuclei,
therefore electrons are mostly captured by the nuclei from the LMP
pool.  As the collapse progresses, the nuclei become heavier and more
neutron rich. Nuclei from the SMMC+RPA pool dominate the rates at
densities from $5 \times 10^{10}$~\gcmic\ to neutrino trapping around
$5 \times 10^{11}$~\gcmic. At even higher densities nuclei from the
FD+RPA pool dominate the rates.

The FD+RPA approach is considered as an approximation to SMMC+RPA
calculations.  To verify its validity we previously compared the
FD-predicted occupation numbers to those obtained from the SMMC
calculation and found fair agreement (Fig.~\ref{fig-Occs}).
The SMMC occupation numbers show some deviation from the
smooth FD behavior, caused by correlations introduced by the
residual interaction, and not reproduced by the FD
distribution.  These deviations translate into differences of the
rates for individual nuclei (see Fig.~\ref{fig-IndRates}).  However,
they do not represent a systematic effect and are smeared out when
the averaging over many nuclei takes place.  This is demonstrated in
Fig.~\ref{fig-AvgXSect-diffBeta} where we compare electron capture
cross sections as derived in the SMMC+RPA approach with those obtained
in the FD+RPA approximation.  The comparison is made for the $250$
nuclei, for which SMMC+RPA rates are now available, adopting for the
NSE averaging the stellar conditions of the $25\Msol$ progenitor
star (Table~\ref{tbl-Zones}). We note that the SMMC pool only includes
heavy nuclei with charge numbers $Z \geq28$, while the capture on the
lighter $pf$ shell nuclei has been evaluated on the basis of the
diagonalization shell model and is here summarized within the LMP
pool.  Fig.~\ref{fig-AvgXSect-diffBeta} demonstrates the results for the
cross section comparison at three distinct points along the stellar
trajectory.  The left plot corresponds to a rather early stage of the
collapse (presupernova phase, index 4) where the SMMC pool covers
$81\%$ of the total abundance of heavy nuclei with $Z\geq 28$.  The
center plot corresponds to collapse conditions just before neutrino
trapping sets in (index 10). Here the $250$ nuclei of the SMMC pool
contribute $99\%$ of the abundance of heavy nuclei with $Z\geq28$ and
dominate the capture rate in the combined pool. The right plot
represents conditions after neutrino trapping (index 13).  The
coverage of the heavy nuclei abundance by the SMMC pool has now
decreased to $60\%$ and the NSE-averaged cross-sections are dominated
by the most neutron-rich nuclei included in the pool.  The FD+RPA
cross sections have been calculated for four different values of the
parameter $\TFD$. The cross sections at fixed electron energy grow
with increasing $\TFD$ as more nucleons are excited across the $N=40$
shell gap opening neutron holes in the $pf$ shell and allowing for
proton excitations within the $gds$ shell.  Differences between the
calculations with different values of $\TFD$ are most noticeable at low
electron energies where the capture process is sensitive to the
details of the GT strength distribution. In turn, the agreement
between the calculations improves for growing electron energy.  We
find that the SMMC+RPA results are best reproduced by the NSE-averaged
FD+RPA cross sections for $\TFD=1.67$ MeV.  We will use in the
following this value for $\TFD$.  Fig.~\ref{fig-AvgRate-diffBeta}
shows that the FD+RPA approach with $\TFD=1.67$ MeV indeed describes
the NSE-averaged SMMC electron capture rates quite well for chemical
potentials $\mue >10$ MeV.  This is the regime of interest as we will use 
the FD+RPA approach to estimate the capture rates for neutron-rich
nuclei which only contribute significantly to the total NSE abundance at
high densities, i.e.\ at the conditions for which the electron chemical
potential is larger than $30$~MeV.

The discussion above took advantage of the 
extended SMMC+RPA pool of $250$ nuclei, which is noticeably larger than
the one adopted in \cite{Langanke03,Hix03}. This increase in coverage
lowered the NSE-averaged rates \cite{Juodagalvis08} at higher values of
electron chemical potential as many of the added nuclei are more
neutron-rich than those used in the original LMSH rates
\cite{Langanke03}.  As explained above, an increase in the neutron
excess usually leads to larger Pauli blocking of GT transitions and
hence smaller rates. Furthermore, more neutron-rich nuclei have larger
$Q$ values which also reduces the electron capture rate.  

Finally we compare electron capture rates for different pools of
nuclei in Fig.~\ref{fig-AvgRate}. The $sd$+LMP+SMMC+FD pool (``all
pools'' in the figure legend) considers around $2700$ nuclei,
including more than $2200$ nuclei for which the rates have been
derived using the FD+RPA approach. The $sd$+LMP+SMMC pool (``no FD''
in the legend) leaves out the FD+RPA rates. A
comparison between the NSE averaged rates for these pools allows an
estimate of the relevance of the  omission of the heavy and most
neutron-rich nuclei for stellar electron capture.  The rates have been
calculated for the conditions along the collapse trajectories of the
$15\Msol$ and $25\Msol$ progenitor stars as listed in
Table~\ref{tbl-Zones}. (The obtained rates are also given in this
table.) As can be observed in Fig.~\ref{fig-AvgRate} the additional
inclusion of the $2200$ heavy and neutron-rich nuclei lowers the
NSE-averaged rates at larger electron chemical potential slightly,
where the added nuclei dominate. 

Fig.~\ref{fig-AvgRate} also quantitatively
demonstrates the impact of the medium effects on the NSE-averaged
rate. As mentioned above, screening effects on the electron capture rates 
lead to a reduction of the
electron capture rates which can amount to almost a factor of $2$ at
large densities (large chemical potentials in Fig.~\ref{fig-AvgRate}).
This is in addition to the effects of screening on the strong and electromagnetic rates that determine the NSE \cite{HT96,Bravo.Garcia-Senz:1999}, which 
is included in \cite{Langanke03,Hix03}.  
Despite this modest reduction
in the rates electron capture on nuclei still dominates over capture
on free protons during the stellar collapse. 

As mentioned in the first section, a very important quantity for
supernova simulations is the spectral distribution of the emitted
neutrinos which becomes relevant especially at higher densities when
neutrinos are becoming trapped. Fig.~\ref{fig-AvgSpectra} illustrates
the calculated spectra of the neutrinos emitted at different stages of
the collapse using the conditions defined by indices 10 and 15 of the
collapse trajectories of the $15\Msol$ and $25\Msol$ progenitor stars
(Table~\ref{tbl-Zones}).  The neutrino spectra are only slightly
changed if we increase the pool of nuclei by more than $2200$ heavy
and neutron-rich nuclei. At the beginning of the collapse these nuclei
do not contribute (this is especially true for the snapshot number 10
of the $25\Msol$ star, where the lines ``full'' and ``no FD'' are
indistinguishable in Fig.\ \ref{fig-AvgSpectra}).  At
$\mue\approx20$~MeV, the SMMC+RPA nuclei dominate.  For the later
collapse phase the inclusion of the FD+RPA nuclei slightly reduces the
average neutrino energy, in agreement with the fact that the heavy
neutron-rich nuclei have slightly larger $Q$ values.

\begin{figure}[tbh]
\begin{center}
\includegraphics*[width=\figsize, angle=0]{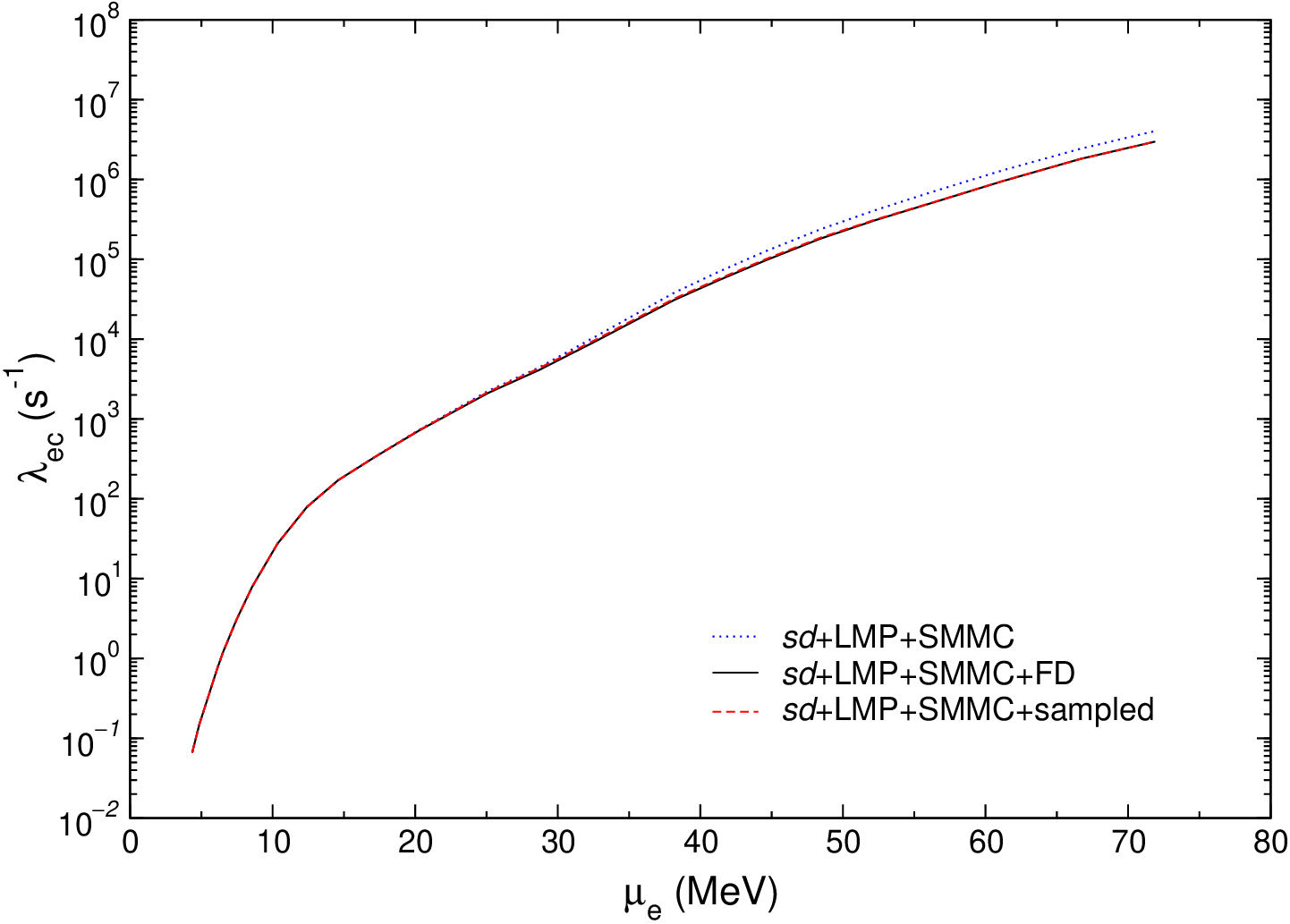}
\caption{ \label{fig-Systematics} (Color online)
Pool-averaged electron capture rates for the trajectory of 
the $25\Msol$ progenitor star.
The rates based on the combined $sd$, LMP and the SMMC+RPA pool
is shown by the dotted line. The averaged rate obtained when
the FD+RPA pool of rates is also included is shown by the solid line.
The dashed line shows the rates if the SMMC-sampled $(Q,\lambda)$
rates (see text) are used to replace the FD+RPA results.
}
\end{center}
\end{figure}

Already during the construction of the LMSH
tabulation~\cite{Langanke03} it was realized that the shell model
approaches can provide rates only for a sample of the nuclei present
in the medium of a collapsing star.  For nuclei not included in the
sample, it was suggested to use a formula based on phase space
considerations with parameters fixed to approximately reproduce the
rates of nuclei in the pool (eq.~(1) in \cite{Langanke03}). A more
careful study of this proposal revealed a deficiency in this approach
\cite{Juodagalvis08}.  For electron chemical potential above
$\mue>40$~MeV, forbidden transitions contribute significantly, and the
value of the ``typical'' transition matrix element $B$, derived for
pure allowed GT transitions, becomes too small. We do not suggest an
alternative parameterization of the rates here.  However, we would
like to exploit the observation that electron capture rates are
becoming insensitive to the detailed structure of the nuclear
transitions with increasing chemical potential and are then relatively
simple functions of $Q$ for a fixed $\mue$.  Using the $250$ nuclei of
the SMMC pool we have derived an (unweighted) average rate for fixed
values of $(Q,\mue)$. Identifying each of the more than $2200$ nuclei,
for which the rate has been calculated solely on the basis of the FD
parametrization, by its $Q$ value, we have derived an alternative set
of capture rates where the rate of each nucleus, as determined by the
FD+RPA approach, has been replaced by the average rate from the SMMC
pool at the corresponding values of $Q$ and $\mu_e$. The two sets of
rates are compared in Fig.~\ref{fig-Systematics}. The agreement
between the sets is excellent, providing confidence that the
limitations of FD+RPA approach are of little impact here.

\section{Conclusions}
\label{sect-Conclusion}

Core-collapse supernova simulations require reliable electron-capture
(and neutrino emission) rates to track the deleptonization process.
In this pursuit we present here an enlarged pool of nuclei from which
we derive NSE-averaged electron capture rates and the corresponding
emitted neutrino spectra.  This extended pool now consists of around
$2700$ nuclei.  The rates of the $sd$ shell nuclei are taken from
Refs.~\cite{Fuller.Fowler.Newman:1980,%
  Fuller.Fowler.Newman:1982b,Fuller.Fowler.Newman:1982a,%
  Fuller.Fowler.Newman:1985}, modified by appropriate screening
corrections.  The $pf$ shell nuclei are adopted from the LMP
rates~\cite{Langanke.Martinez-Pinedo:2001}. The SMMC+RPA set
\cite{LMS,Langanke03} was enlarged to include in total $250$
nuclei~\cite{Juodagalvis08}. Additionally, more than $2200$ nuclei are
calculated within the FD+RPA approach, adjusted to reproduce the
SMMC+RPA rates. In the FD+RPA approach the parent nucleus occupation
numbers at finite temperature are approximated by a Fermi-Dirac
parameterization taking the parameter value $\TFD=1.67$ MeV.  The
latter three sets of rates include screening effects directly, while
for the FFN rates we use an approximate, but rather accurate,
prescription. This huge number of individual electron-capture rates
was used to obtain the averaged electron-capture rates at various
stellar conditions $(\rho,T,Y_e)$. The resulting table is available
upon request from the authors.

\begin{ack}

  At the early stage of the project AJ's work was partly supported by
  the US DOE through the SciDAC programme ``Terascale Supernova
  Initiative'' during his stay at GSI, Germany, hospitality of which
  is appreciated. AJ acknowledges support from the EU FP6-FP7 project
  ``BalticGrid''.  GMP and KL acknowledge support by the Deutsche
  Forschungsgemeinschaft through contract SFB 634, by the ExtreMe
  Matter Institute EMMI and by the Helmholtz International Center for
  FAIR. WRH acknowledges support from the Office of Nuclear Physics,
  US Department of Energy.  Discussions with Eduardo Bravo about the
  treatment of screening corrections are acknowledged.  The
  computer-net {\sf BalticGrid} infrastructure was used to calculate
  this huge number of rates and spectra.

\end{ack}


\appendix*

\section{Screening corrections to electron capture rates}
\label{sec:scre-corr-electry}

Coulomb corrections are known to play an important role in
determining the thermodynamical properties of a high density
plasma~\cite{Yakovlev.Shalybkov:1989}. At the conditions we are
interested in this work the matter can be assumed to be in nuclear statistical
equilibrium. This regime has been studied
by Bravo and Garc{\'i}a-Senz~\cite{Bravo.Garcia-Senz:1999} and we have generalized their
treatment to the study of electron capture rates. When Coulomb
corrections are included the chemical potential of the nuclear species
is given by:
\begin{equation}
  \label{eq:1}
  \mu_i = \mu_{i,0} + \mu_{i,C}.
\end{equation}
Here $\mu_{i,0}$ is the chemical potential in the absence of Coulomb
effects which is generally given by the Boltzmann statistics, and
$\mu_{i,C}$ is the contribution to the chemical potential due to the
interaction of nucleus $i$ with the electron background. 

The core of the star constitutes a multicomponent plasma that we will
treat in the additive approximation. In this case, all the
thermodynamic quantities are computed as the sum of the individual
quantities for each species. If one further assumes that the
electron distribution is not affected by the presence of the nuclear
charges (uniform background approximation), the Coulomb chemical
potential of species $i$ is given by~\cite{Yakovlev.Shalybkov:1989}
\begin{equation}
  \label{eq:2}
  \mu_{i,C} = k_B T f_C(\Gamma_i),
\end{equation}
where $f_C$ is the Coulomb free energy per ion in units of $k_B T$ and
$\Gamma_i$ is the ion-coupling parameter,
\begin{equation}
  \label{eq:3}
  \Gamma_i = Z_i^{5/3} \Gamma_e = \frac{Z_i^{5/3} e^2}{a_e k_B T},
\end{equation}
where $a_e$ is the electron sphere radius, $a_e = (3/(4\pi
n_e))^{1/3}$, with $n_e$ the electron density.

For the free energy in the regime $\Gamma > 1$, we use the
expression~\cite{Yakovlev.Shalybkov:1989,Slattery.Doolen.Dewitt:1982}: 
\begin{equation}
  \label{eq:4}
  f_C(\Gamma) = a \Gamma + 4 b \Gamma^{1/4} - 4 c \Gamma^{-1/4} + d
  \ln \Gamma + e,
\end{equation}
with the values of the parameters $a, b, c, d$ and $e$
taken from~\cite{Ichimaru:1993}:
\begin{eqnarray}
  \label{eq:5}
  a &=& -0.898004, \quad b = 0.96786, \quad c = 0.220703\nonumber\\
  d &=& -0.86097, \quad e = -2.52692.
\end{eqnarray}

For $\Gamma < 1$, we use the expression suggested in
ref.~\cite{Yakovlev.Shalybkov:1989}: 
\begin{equation}
  \label{eq:6}
  f_C(\Gamma) = -\frac{1}{\sqrt{3}} \Gamma^{3/2} +
  \frac{\beta}{\gamma} \Gamma^\gamma. 
\end{equation}

The first term reproduces the Debye-H\"uckel limit for $\Gamma\ll 1$
and the parameters $\beta$ and $\gamma$ are determined requiring the
continuity of the internal energy and its derivative 
at $\Gamma = 1$. These conditions give $\beta = 0.295614$ and $\gamma =
1.98848$.

The Coulomb corrections contribute to the electron capture rates in
two different ways. First, due to the fact that the chemical potential
depends on $Z$, screening effects will change the threshold energy for
the capture by the amount~\cite{Couch.Loumos:1974}:
\begin{equation}
  \label{eq:7}
  \Delta Q_C = \mu_C(Z-1) - \mu_C(Z),
\end{equation}
where $Z$ is the charge number of the capturing nucleus.  As $\mu_C$
is negative, this correction increases the energy threshold and thus
reduces the electron capture rate.

Second, the energy of the captured electron is affected by the
presence of the background electron gas. Its energy will be reduced
compared to the unscreened case. The
magnitude of this effect can be determined using linear response
theory~\cite{Itoh.Tomizawa.ea:2002}. Moreover, we can assume that the
screening potential, $V_s$, is constant inside the nucleus
with a value which can be determined from equation (17)
of Itoh et al.~\cite{Itoh.Tomizawa.ea:2002} evaluated at the nuclear radius.

To consider both screening corrections the phase space integral
\cite{Langanke.Martinez-Pinedo:2000} gets modified. 
The rate of the electron capture during the nuclear transition from 
an initial state $i$ to a final state $f$~\footnote{Note that here the
  state denotes
  either a single particle state if the rates are determined using an
  RPA approach as in the present calculations or a many body state if
  the shell-model or similar approach is used.}
is given by:
\begin{equation}
  \label{eq:ratescr}
  \lambda^{\rm ec}_{i\!f} = \frac{1}{\pi^2\hbar^3}
   \int_{\varepsilon_e^{0s}}^{\infty} p_e^2
  \sigmaEC(\varepsilon_e,\varepsilon_i,\varepsilon_f)
  f(\varepsilon_e+V_s,\mu_e,T) d\varepsilon_e,
\end{equation}
$\varepsilon_e^{0s}=\max(\Qif^s,m_e c^2)$, and
the energy of the emitted neutrino is related to the electron
energy by $\varepsilon_\nu = \varepsilon_e - \Qif^s$, where
\begin{equation}
  \label{eq:9}
  \Qif^s = \Qif + \Delta Q_C,
\end{equation}
with $\Qif$ being the capture threshold in the absence of
screening corrections. 

Assuming a Fermi-Dirac distribution for the electrons and using the
property $f(\varepsilon_e+V_s,\mu_e,T)=f(\varepsilon_e,\mu_e-V_s,T)$
one can rewrite equation~(\ref{eq:ratescr}) as follows:
\begin{equation}
  \label{eq:ratescr2}
  \lambda^{\rm ec}_{i\!f} = \frac{1}{\pi^2\hbar^3}
  \int_{\varepsilon_e^{0s}}^{\infty} p_e^2
  \sigmaEC(\varepsilon_e,\varepsilon_i,\varepsilon_f)
  f(\varepsilon_e,\mu_e-V_s,T) d\varepsilon_e.
\end{equation}
The presence of the electron background effectively reduces the
electron chemical potential and hence also reduces the capture rate.
In summary, in order to evaluate the electron capture rates under the
presence of screening the threshold energy should be modified
following equation~(\ref{eq:9}) and the chemical potential of the
electrons should be reduced by $V_s$.

The formalism discussed above allows to include the screening
corrections in the evaluation of electron capture rates provided that
one uses some nuclear model to determine the initial and final states.
However, many currently available tabulations of electron capture
rates~\cite{%
  Fuller.Fowler.Newman:1980,%
  Fuller.Fowler.Newman:1982b,Fuller.Fowler.Newman:1982a,%
  Fuller.Fowler.Newman:1985,%
  Langanke.Martinez-Pinedo:2001,%
  Oda.Hino.ea:1994,%
  Nabi.Klapdor-Kleingrothaus:1999a,%
  Pruet.Fuller:2003}
have been computed without including
screening corrections. (Note that for this paper we recalculated the
LMP rates \cite{Langanke.Martinez-Pinedo:2001} 
with the screening corrections included directly.)
For such tabulations it is thus important to
determine an aposteriori prescription to incorporate screening effects
for the use of such rate tables in astrophysical simulations.

\begin{figure}[b]
  \centering
  \includegraphics[width=\figsize]{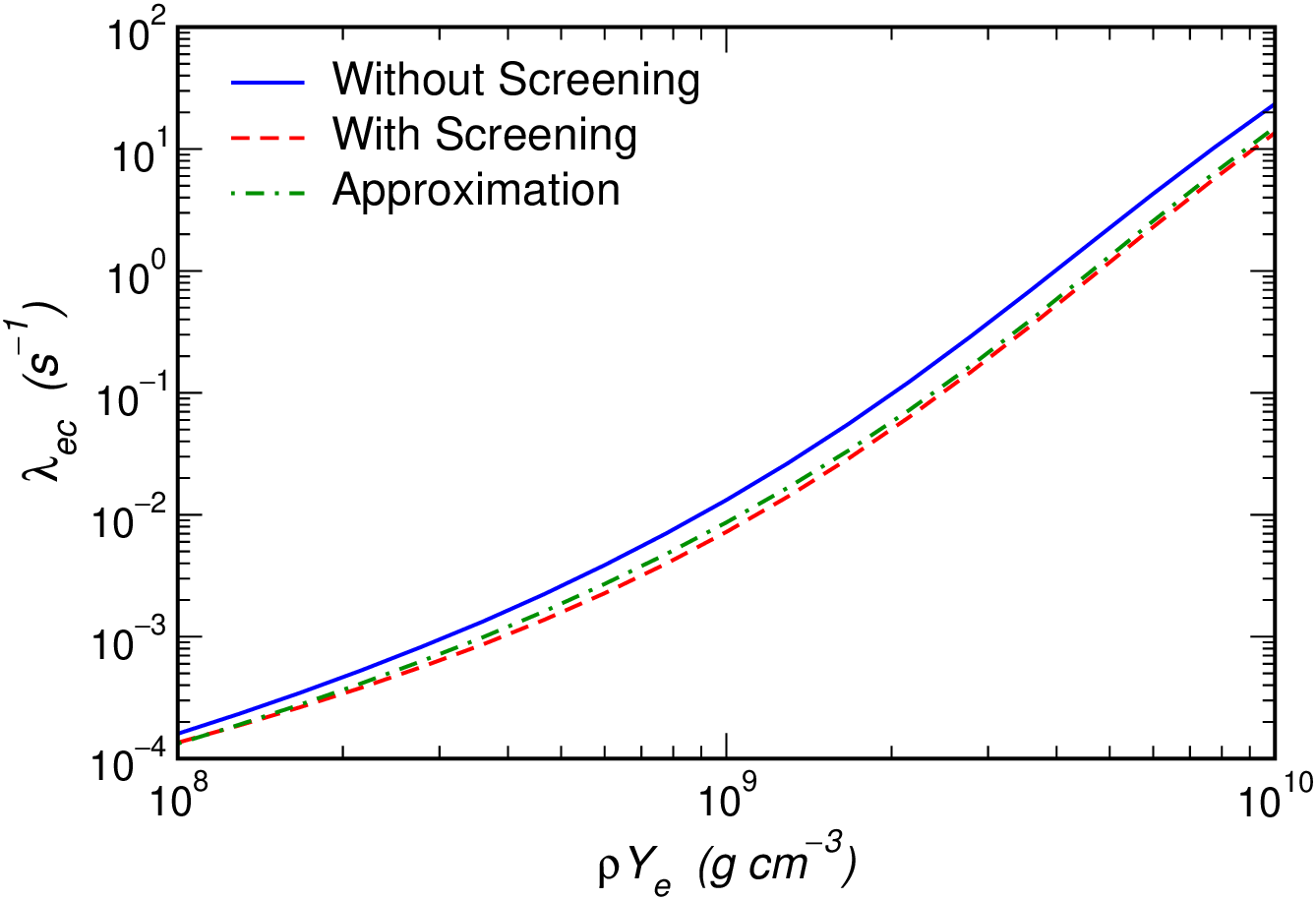}
  \caption{Rate for electron capture on $^{58}$Fe for a temperature of
    7.5~GK as a function of the electron density, $\rho Y_e$. The
    solid line corresponds to the rate computed in
    ref.~\cite{Langanke.Martinez-Pinedo:2001}, the dashed line uses
    the same matrix elements but includes screening corrections as
    discussed in the text, see eq.~(\ref{eq:ratescr2}). The
    dashed-dotted line has been obtained using the approximation
    suggested in equation~(\ref{eq:13}).\label{fig:fe58rat}}
\end{figure}

A very good approximation can be obtained by using the effective $ft$
value formalism introduced by Fuller et al.~\cite{Fuller.Fowler.Newman:1985} which
allows accurate interpolations of the tabulated electron capture
rates.  In this approach an effective $ft$-value is defined that
represents an average nuclear matrix element which characterizes the
capture rate on a given nucleus:
\begin{equation}
  \label{eq:11}
  ft_{\rm eff} \equiv \ln 2 \frac{F(Q,\mu_e)}{\lambda_{\rm ec}},
\end{equation}
where $\lambda_{\rm ec}$ is the tabulated electron capture rate,
$Q$ is the ground-state to ground-state transition energy, and
$F$ is the integral
\begin{equation}
  \label{eq:12}
  F(Q,\mu_e) = \frac{1}{m_e^5 c^{10}} \int_{\max(Q,m_e c^2)}^\infty w^2 (Q
  + w)^2 f(w,\mu_e,T) dw,  
\end{equation}
Assuming that the average nuclear matrix element $ft_{\rm eff}$ is
unaffected by screening, the ratio of electron capture rates with and
without screening is equal to the ratio of $f$ functions, leading to:
\begin{equation}
  \label{eq:13}
  \lambda_{\rm ec}^C = \frac{F(Q+\Delta
    Q_C,\mu_e - V_s)}{F(Q,\mu_e)} \lambda_{\rm ec}  
\end{equation}

Figure~\ref{fig:fe58rat} compares the electron capture rate on
$^{58}$Fe computed using the shell-model calculations
of~\cite{Langanke.Martinez-Pinedo:2001} with and without screening
corrections. In addition, the approximation~(\ref{eq:13}) is also
shown. The difference between the exact implementation of screening
corrections, eq.~(\ref{eq:ratescr2}), and the approximation is never
larger than $20\%$ for the conditions shown in figure~\ref{fig:fe58rat}.


\end{document}